\documentclass[DIV=11]{scrartcl}
\usepackage[utf8]{inputenc}

\let\latextextsuperscript\textsuperscript
\AtBeginDocument{\let\textsuperscript\latextextsuperscript}

\usepackage[colorlinks,draft=false]{hyperref}

\usepackage{authblk}
\author[1]{Arjan Cornelissen}
\author[2]{Johannes Bausch}
\author[3]{András Gilyén}
\affil[1]{QuSoft, University of Amsterdam} 
\affil[2]{CQIF, DAMTP, University of Cambridge}
\affil[3]{IQIM, California Institute of Technology}

\newcommand{\hidepics}{false}

\usepackage{floatpag}
\newcommand{\ExternalLink}{%
    \tikz[x=1.2ex, y=1.2ex, baseline=-0.05ex]{%
        \begin{scope}[x=1ex, y=1ex]
            \clip (-0.1,-0.1) 
                --++ (-0, 1.2) 
                --++ (0.6, 0) 
                --++ (0, -0.6) 
                --++ (0.6, 0) 
                --++ (0, -1);
            \path[draw, black,
                line width = 0.5, 
                rounded corners=0.5] 
                (0,0) rectangle (1,1);
        \end{scope}
        \path[draw, black, line width = 0.5] (0.5, 0.5) 
            -- (1, 1);
        \path[draw, black ,line width = 0.5] (0.6, 1) 
            -- (1, 1) -- (1, 0.6);
        }
    }

\usepackage{amsthm,amsmath,amssymb,mathtools}
\usepackage{microtype,dsfont,newtxtext,newtxmath,xfp}
\usepackage{mleftright}\mleftright
\usepackage{graphicx,tikz}
\usepackage{grffile}
\usepackage[hang,flushmargin]{footmisc} 
\usetikzlibrary{backgrounds,scopes,quantikz}
\usepackage{subcaption,booktabs,multirow,makecell,siunitx,array}
\newcolumntype{L}[1]{>{\raggedright\arraybackslash}p{#1}}
\newcolumntype{R}[1]{>{\raggedleft\arraybackslash}p{#1}}

\usepackage[section]{placeins}
\makeatletter
\AtBeginDocument{%
  \expandafter\renewcommand\expandafter\subsection\expandafter{%
    \expandafter\@fb@secFB\subsection
  }%
}
\makeatother

\input{braket}

\usepackage{ifdraft,verbatim}
\ifdraft{\usepackage{showlabels}}{}
\ifdraft{\newcommand{\authnote}[3]{{\color{#3} {\textbf{ #1:}} #2}}}{\newcommand{\authnote}[3]{}}

\newcommand\op\mathbf
\newcommand\field\mathds
\newcommand\ee{\mathrm e}
\newcommand\ii{\mathrm i}

\DeclareMathOperator{\Prob}{Prob}
\DeclareMathOperator{\CBell}{C_\mathrm{Bell}}
\DeclareMathOperator{\argmax}{argmax}

\usepackage[capitalise]{cleveref}
\Crefname{lemma}{Lemma}{Lemmas}
\Crefname{proposition}{Proposition}{Propositions}
\Crefname{definition}{Definition}{Definitions}
\Crefname{theorem}{Theorem}{Theorems}
\Crefname{conjecture}{Conjecture}{Conjectures}
\Crefname{corollary}{Corollary}{Corollaries}
\Crefname{example}{Example}{Examples}
\Crefname{section}{Section}{Sections}
\Crefname{appendix}{Appendix}{Appendices}
\Crefname{figure}{Fig.}{Figs.}
\Crefname{equation}{Eq.}{Eqs.}
\Crefname{table}{Table}{Tables}
\Crefname{item}{Property}{Properties}
\Crefname{remark}{Remark}{Remarks}

\usepackage[backend=biber,style=alphabetic,sorting=anyt,doi=false,isbn=false,url=false,maxbibnames=20,maxcitenames=2]{biblatex}
\bibliography{bib-andras,bib-arjan,bib-jo,LocalBibliography}
\newbibmacro{string+doi}[1]{\iffieldundef{doi}{#1}{\href{https://dx.doi.org/\thefield{doi}}{#1}}}
\DeclareFieldFormat{title}{\usebibmacro{string+doi}{\mkbibemph{#1}}}
\DeclareFieldFormat[article]{title}{\usebibmacro{string+doi}{\mkbibquote{#1}}}
\DeclareFieldFormat[incollection]{title}{\usebibmacro{string+doi}{\mkbibquote{#1}}}

\title{Scalable Benchmarks \\ for Gate-Based Quantum Computers}

\date{April 20, 2021}

\begin{document}
\maketitle

\begin{abstract}
    In the near-term ``NISQ''-era of noisy, intermediate-scale, quantum hardware and beyond, reliably determining the quality of quantum devices becomes increasingly important: users need to be able to compare them with one another, and make an estimate whether they are capable of performing a given task ahead of time. 
	In this work, we develop and release an advanced quantum benchmarking framework in order to help assess the state of the art of current quantum devices.
	
	Our testing framework measures the performance of universal quantum devices in a hardware-agnostic way, with metrics that are aimed to facilitate an intuitive understanding of which device is likely to outperform others on a given task.
	This is achieved through six structured tests that allow for an immediate, visual assessment of how devices compare.
	Each test is designed with scalability in mind, making this framework not only suitable for testing the performance of present-day quantum devices, but also of those released in the foreseeable future.
	The series of tests are motivated by real-life scenarios, and therefore emphasise the interplay between various relevant characteristics of quantum devices, such as qubit count, connectivity, and gate and measurement fidelity.
	
	We present the benchmark results of twenty-one different quantum devices from IBM, Rigetti and IonQ.
\end{abstract}

\clearpage

\tableofcontents

\section{Why Quantum Benchmarks Matter}
When buying a new laptop, we hope that the new machine will be better than the old one.
``Better'', of course, will largely depend on who buys the new machine: a subjective mix of ``larger memory'', ``faster CPU'', or ``spiffier color'' will decide the fate of which model is bought, and which one will collect dust on the shelf.
But, color aside, how do we compare how fast a computer is?

Strengths and weaknesses of different computer hardware or hardware combinations are commonly assessed via benchmarks that test various aspects and return a score that allows for a quantitative comparison.
Most commonly, the metrics we measure are those of performance: how \emph{fast}---and increasingly important how \emph{efficient}---is the computer? Will the combination of CPU and GPU be able to run the next-gen game, will the memory suffice for the database application, and will the storage be swift enough for HD video editing?
    
Such benchmarks have a long history for classical PC hardware, since they enable the end-user to compare devices independently, beyond the data and specification sheet that vendors provide.
A well-known example was the race for ever higher CPU frequencies by Intel in the early 2000's, which AMD countered with model names like ``AMD Athlon 3500+``, suggesting a ``comparative performance'' of a 3500MHz Intel CPU, even though their devices were clocked at a much slower speed.
This exemplifies that raw device parameters fall short of delivering the full picture of what a computer is capable of; a benchmark is necessary to reliably assess the device's performance.

While this race for raw computing power has continued for decades in the case of classical computers, it is also starting to pick up pace on the quantum side.
Headlines announce ever larger quantum devices as more and more companies join the race for the most usable quantum computers.
But how do we compare these early-stage quantum computers, if not even the architectures match---if one device features superconducting qubits, and another one shuttles ions around in a magnetic trap?
With companies' rising interest in quantum technologies, how can they choose a device that will be able to run a circuit that will solve the company's next business case?\footnote{Spoiler alert: Bitcoin mining is not yet possible.}

\paragraph{A Brief History of Quantum Benchmarking.}

We are only in the early days of quantum computing, featuring near-term noisy intermediate-scale quantum devices with very restricted capabilities---with approximately $50$ qubits and the ability to run circuits with depths of around $20$ \cite{Google2019}.
While the term ``quantum benchmarking'' is widely used today, it is related only by virtue of testing and tuning device fidelity on a device-by-device basis---often using random circuits in the context of randomized benchmarking \cite{Emerson2005}---yet with methods that vary from vendor to vendor; a comparative metric is not generally extracted.
Furthermore, one expects random circuits to, in expectation, blur out systematic errors present in the device.
Structured circuits (which are the circuits one commonly aspires to execute on a quantum computer) are thus much more sensitive to device noise; a fact that has recently been verified experimentally \cite{Proctor2020}.

Over the past years, interest in comparing the performance of quantum computers has grown significantly, and various structured tasks have been proposed within which such a comparison is possible, such as building blocks of digital quantum circuits \cite{Linke2017}, preparation of absolute maximally entangled states \cite{Cervera-Lierta2019}, quantum chemistry and material science simulations \cite{Bauer2020a,McCaskey2019}, or random variational circuits \cite{Mills2020}, just to name a few.
In 2018, IBM proposed a single-value metric (based on \cite{Moll2018}) for quantum devices, called ``quantum volume'' \cite{Cross2019}; it quantifies how large a random circuit with circuit depth equal to qubit grid size (i.e.\ a ``square circuit''), a device is capable of executing with reasonable fidelity, via the formula $\log_2 V_Q=\argmax_m \{\min [m, d(m)]\}$,
where $m$ is the sidelength of the square randomized circuit, and $d(m)$ is the maximum depth such that said circuit still produces a heavy-tailed distribution. Assuming all-to-all connectivity, one can approximate $d(m) \approx 1/m\epsilon$, where $\epsilon$ is the average two-qubit gate error rate between any pairs; for a linear topology, $d(m) \approx 1/m^2 \epsilon$.
Quantum volume has been shown to correlate with the quality of a device within various contexts, such as variational optimization of quantum chemistry applications~\cite{Moll2018}; at the time of writing of this version of the paper, state-of-the-art devices feature a quantum volume of 64 \cite{Jurcevic2020}, and larger ones have recently been announced.
A more recent suggestion by Atos is called \emph{Q-score}, which counts the number of variables in a max-cut problem that a device can optimize; current Q-scores range between 10 and 20 \cite{Atos-QScore}.

Yet such one-dimensional measures appear overly simplistic to completely characterise a quantum device's performance; even for personal computers a plethora of metrics is commonly provided, and both synthetic and real-world benchmarks are executed that test various aspects about a device (such as GPU memory transfer speeds, floating point performance, cache latency, or texture fill rates).
Even if those individual scores are then collated into a final ``meta-score'', we would expect this score to be more reliable than a one-dimensional assessment can be.

\paragraph{Our Contribution.}
We created a quantum benchmark suite in order to provide a comparison tool for the currently available and upcoming quantum computing platforms from an end-user perspective. Our goal is to help understanding the performance of the available devices via real-world-inspired use cases, by providing meaningful benchmark scores for a series of different tests. We present benchmarks tailored to the small quantum devices currently available.
As near-term quantum devices are relatively limited in their capabilities, we place emphasis on test scenarios that can also discriminate performance across devices that lie on the lower-performance end of the NISQ spectrum: all our metrics are readily available for devices with two qubits only and circuit depths $<10$. However, we keep scalability in mind (see \cref{sec:scalability}), and for all benchmarks it should be possible to execute tweaked variants on larger devices when they become available.

The benchmarks we chose all feature an easily-depictable representation of the outcome, which demonstrates unambiguously, and at a single glance, which device holds the lead in a direct comparison: this allows us to delineate performance gains for newer quantum hardware qualitatively, and follow the progress of device manfacturers' timelines.
Moreover, the visualizations sometimes reveal interesting features of the noise, such as fluctuations and changes of device calibration over time, as well as other systematic errors that result in deformed images.
This emphasis on presenting the benchmark results \emph{visually} is much akin to how graphics hardware is often tested by how detailed a rendered image looks.

Nonetheless, we provide a numerical measure including uncertainties for each of the benchmarks; these numerical metrics characterise different noise aspects in a more precise manner than a visual depiction can. Furthermore, numerical scores allow us to compare our metrics to others, for instance to quantum volume (see \cref{sec:qv}), and we find a strong correlation between these scores.

The circuits corresponding to the benchmarks are all implemented using several vendors' SDKs and APIs 
(Qiskit / IBMQ for IBM, Forest / QCS and Amazon Braket for Rigetti, Amazon Braket for IonQ, and cirq for Google),
enabling us to execute the tests on various quantum devices available to us. Our python code is bundled into a simple command-line application, which allows for spawning the tests, managing the communication with the vendor's API's, as well as post-processing the results and rendering the visualizations. All of this is packed into an open-source Github repository~\cite{github-repo}, freely available to be used and re-used by the community.

\paragraph{Paper Overview.}
In \cref{sec:benchmarks}, we describe on a theoretical level each of the six structured circuit families that we implement.
The benchmark results---based on over one billion circuit runs combined---are presented in \cref{sec:results}, alongside an extensive discussion of circuit depths, gate counts, and error metrics employed.

\section{Structured Circuits as Visual Benchmarks}\label{sec:benchmarks}

We developed six benchmarks, all of which test slightly different aspects of the quantum hardware; the tests and the emphasis they place on the hardware are listed in \cref{tab:benchmarks}.
Naturally, all tests jointly test gate fidelity, readout noise, and the ability of the compilers to take full advantage of the underlying device topology.

\begin{table}[!tb]
	\small\centering
	\begin{tabular}{R{4cm}L{4cm}L{2.3cm}}
		\toprule
		Benchmark & Main tested aspect & Visualization \\
		\midrule
		Schrödinger's Microscope Mandelbrot & wide shallow circuits: postselection, routing & fractal \\[9mm]
		Line Drawing & \makecell[l]{deep narrow circuits:\\state preparation, QFT,\\tomography} & figure outline \\[9mm]
		Platonic Fractals & weak deferred measurements & L-system \\[5mm]
		Gate-efficient HHL & QSVT, matrix inversion (quantum linear algebra) & histogram \\[9mm]
		Bell Test & linear circuits: entanglement capability, coherence range & heatmap \\
		\bottomrule
	\end{tabular}
	\caption{Overview of the implemented benchmarks, which device aspect they test in particular, and what the resulting visualization looks like. All benchmarks can be run on devices that have at least $2$ qubits; the Schr\"odinger's Microscope and Mandelbrot test yield circuits on a number of qubits that is a power of $2$.}
	\label{tab:benchmarks}
\end{table}

\subsection{Bell Test}

\begin{figure}[!t]
\centering
\subcaptionbox{%
A hypothetical qubit interaction graph.
\label{fig:bell-topology}
}[.44\linewidth]{%
\centering
\begin{tikzpicture}[
    qubit/.style={
        circle,
        minimum size=5mm,
        draw=none,
        fill=black!20
    }
]
\foreach \x/\y/\tag in {-.41/0/Q0,1/0/Q1,2/1/Q2,3/0/Q3,2/-1/Q4,4.41/0/Q5}{
    \node[qubit] (\tag) at  (\x,\y) {\tag};
}
{[on background layer]
\draw (Q0)--(Q1)--(Q2)--(Q3)--(Q4)--(Q1) (Q3)--(Q5);
}
\end{tikzpicture}
}
\\[.8cm]
\subcaptionbox{%
Bell test between qubits Q0 and Q3, where we always choose the path that has the highest sum of gate fidelities for the above circuit, as provided by the vendor (e.g.\ as shown here, Q0--Q1--Q2--Q3).
Both $\op R_A$ and $\op R_B$ are $\op Z$ rotations with angles $(\theta_A,\theta_B) \in \{ (0, \pi/3), (0, 2\pi/3), (\pi/3, 2\pi/3) \}$.
The measurements are performed in the standard basis.
\label{fig:bell-circuit}
}[.8\linewidth]{%
\centering
\begin{quantikz}[row sep={0.75cm,between origins},column sep=.5cm]
\lstick{$\ket{0}_\mathrm{Q0}$} & \gate{\op X} & \gate{\op H} & \ctrl1  & \qw     & \qw     & \qw     & \ctrl1\slice{$\ket{\phi_1}$}  & \gate{\op R_A} & \gate{\op  H}\slice{$\ket{\phi_2}$} & \meter{} \\
\lstick{$\ket{0}_\mathrm{Q1}$} & \qw     & \qw     & \targ{} & \ctrl1  & \qw     & \ctrl1  & \targ{} & \qw          & \qw     & \qw \\
\lstick{$\ket{0}_\mathrm{Q2}$} & \qw     & \qw     & \qw     & \targ{} & \ctrl1  & \targ{} & \qw     & \qw          & \qw     & \qw \\
\lstick{$\ket{0}_\mathrm{Q3}$} & \gate{\op X} & \qw     & \qw     & \qw     & \targ{} & \qw     & \qw     & \gate{\op R_B} & \gate{\op H} & \meter{}
\end{quantikz}
}
\caption{Example of a Bell test circuit run for the hypothetical interaction graph in \cref{fig:bell-topology}.}
\label{fig:bell}
\end{figure}

One of the basic tests we propose is a simple Bell test between any pair of qubits in the circuit.
The original Bell inequality \cite[Eq.~15]{Bell1964} reads $|\Prob(\vec a, \vec b) - \Prob(\vec a, \vec c)| \le 1 + \Prob(\vec b, \vec c)$, for three measurement directions $\vec a, \vec b, \vec c$ with associated angles $\theta_a, \theta_b, \theta_c$ (e.g.\ the directions of a beam splitter, and where $\Prob(\vec a, \vec b)$ denotes the likelihood of measuring direction $\vec a$ with the first device, and $\vec b$ with the second).
Within a classical theory without entanglement between the measured entities we would expect the correlation between any two measurements to equal those probabilities, i.e.\ $\Prob(\vec x, \vec y) = C(\theta_x, \theta_y)$.
On the other hand, for a singlet state in quantum mechanics (cf.~\cite[Eq.~3]{Bell1964}), we have that $C(\theta_x, \theta_y) = -\cos(\theta_x - \theta_y)$ is determined by the relative angles between the two respective measurement directions. 
For a choice of $(\theta_a, \theta_b, \theta_c) = (0^{\circ}, 60^{\circ}, 120^{\circ})$, each of the correlation coefficients equals $1/2$ in magnitude; violating the Bell inequality as $1/2 - (-1/2) \not\le 1 - 1/2$.

The specific Bell experiment we propose here is set up as shown in \cref{fig:bell},
where a route along the given topology is chosen that minimizes the gate error (we rely on the one- and two-qubit gate fidelities provided by the vendor).
Along this route, and between start and end point, we create a $\ket-$ Bell pair; i.e., $\ket{\phi_1}$ shown in \cref{fig:bell-circuit} equals $\ket{\phi_1} = \ket{\Psi^-}_\mathrm{Q0,Q3} \ket0_\mathrm{Q1} \ket0_\mathrm{Q2}$, where $\ket{\Psi^-}\coloneqq (\ket{01} - \ket{10})/\sqrt 2$, as can be easily verified.

We then choose two $\op Z$ rotation gates $\op R_A$ and $\op R_B$ with angles $\theta_A$ and $\theta_B$ given below, and then measure in the $\op X$ basis.
Prior to measurement, the state $\ket{\phi_2}$ as shown in \cref{fig:bell-circuit} (where we drop the qubits not measured) is
\begin{equation}\label{eq:bell-before-measurement}
\ket{\phi_2} = \frac{1}{4\sqrt2}\begin{cases}
    \left[ -1+\ii\sqrt3, -3-\ii\sqrt3, 3+\ii\sqrt3, 1-\ii\sqrt3 \right], & (\theta_A,\theta_B) = (0, \pi/3), \\[1mm]
    \left[ -3+\ii\sqrt3, -1-\ii\sqrt3, 1+\ii\sqrt3, 3-\ii\sqrt3 \right], & (\theta_A,\theta_B) = (0, 2\pi/3), \\[1mm]
    \left[ -2, -2\ii\sqrt3, 2\ii\sqrt3, 2 \right], & (\theta_A,\theta_B) = (\pi/3, 2\pi/3).
\end{cases}
\end{equation}
We get the measurement probabilities
\begin{equation}
    p_\mathrm{eq} \coloneqq \Prob(00 \lor 11) = \begin{cases}
    1/4, & (\theta_A,\theta_B) = (0, \pi/3) \lor (\theta_A,\theta_B) = (\pi/3, 2\pi/3), \\
    3/4, & (\theta_A,\theta_B) = (0, 2\pi/3),
    \end{cases}
\end{equation}
and $p_\mathrm{ineq} \coloneqq 1 - p_\mathrm{eq}$.
This gives rise to the correlation coefficients \[
C(\theta_A,\theta_B)=p_\mathrm{eq} - p_\mathrm{ineq} = \begin{cases}
-1/2, & (\theta_A,\theta_B) = (0, \pi/3) \lor (\theta_A,\theta_B) = (\pi/3, 2\pi/3), \\
1/2, & (\theta_A,\theta_B) = (0, 2\pi/3).
\end{cases}
\]
Thus, if the computation is perfect, then $\CBell := C(0,2\pi/3) - C(0,\pi/3) - C(\pi/3,2\pi/3) = 3/2$.
In contrast, if the state $\ket{\phi_1}$ were not entangled, then the Bell inequality must hold: $\CBell \le 1$ \cite{Bell1964}.
Thus, an outcome in the interval $(1,3/2]$ witnesses entanglement between the two qubits, and the closer to $3/2$, the less noise there was present in the device.

As shown in \cref{fig:bell-circuit}, the circuits in this family comprise almost exclusively CNOT gates (apart from five single-qubit gates to transform into the Bell basis). As such, the test is particularly well-suited to assess circuits of a linear nature, where individual gates are executed in a sequential fashion; the circuits are twice as deep as they are wide ($2d+1$ for two qubits of distance $d$ with respect to the device topology).
We remark that the circuits have an implicit \emph{directedness} due to the directed nature of the CNOT gates; we therefore apply the test between any pairs of qubits $(A,B)$ as well as $(B,A)$. This highlights whether CNOTs can be applied in both orientations with comparable fidelity.

\subsection{Benchmarks based on complex transformations of the Riemann sphere}\label{sec:Riemann}

The amplitudes in quantum states are in general complex numbers. 
The gates that one applies in a quantum device modify these complex amplitudes.
Thus, even though the quantum Holevo bound restricts how much information can ever be extracted from these complex amplitudes in a single round of measurements, internally, a quantum device should be able to do computations with complex numbers naturally at a very low level.
The goal of these two benchmarks is to exploit this feat.

We first define~\cite{kiss2006ComplChaosCondQubitDyn} a continuous bijection $z \mapsto \ket{\psi_z}$ from the Riemann sphere $\overline{\mathbb{C}} = \mathbb{C} \cup \{\infty\}$ to the set of single qubit quantum states, as
\[\forall z \in \mathbb{C}, \ket{\psi_z} = \frac{z\ket{0} + \ket{1}}{\sqrt{|z|^2 + 1}} \qquad \text{and} \qquad \ket{\psi_{\infty}} = \ket{0}.\]
Note in particular that any single qubit quantum state $\alpha\ket{0} + \beta\ket{1}$ can be written in this form by setting $z = \alpha/\beta$; and to prove continuity at infinity it is important to remember that quantum states are rays in a Hilbert space, defined only up to a global phase.
Any desired state $\ket{\psi_z}$ can be prepared from a pair of rotations, via
\[\ket{\psi_z} = Z^{\phi_2}Y^{\phi_1}\ket{0} \qquad \text{where} \qquad \phi_1 = \frac{2}{\pi}\arccos\left(\frac{|z|}{\sqrt{|z|^2 + 1}}\right) \qquad \text{and} \qquad \phi_2 = -\frac{\arg(z)}{\pi}.\]

Under this identification between the single qubit quantum states and $\overline{\mathbb{C}}$, single qubit unitary operations induce M\"obius transformations
\[
    z \longmapsto \frac{az + b}{cz + d}
\]
encoded in the amplitudes via
\[U\ket{\psi_z} \propto \begin{bmatrix}
	a & b \\ c & d
\end{bmatrix}\begin{bmatrix}
	z \\ 1
\end{bmatrix} = \begin{bmatrix}
	az + b \\ cz + d
\end{bmatrix} \propto \begin{bmatrix}
	\frac{az + b}{cz + d} \\ 1
\end{bmatrix} \propto \ket{\psi_{\frac{az+b}{cz+d}}},\]
where in steps marked with $\propto$ we neglect normalization.
To see that this map is indeed a well-defined Möbius transformation it is sufficient to note that $U$ is necessarily full rank---so in particular $ad-bc\neq 0$.

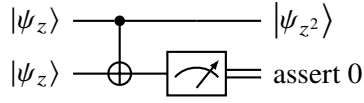
\begin{figure}[!t]
	\centering
	\begin{quantikz}[row sep={0.7cm,between origins},column sep=.45cm]
		\lstick{$\ket{\psi_z}$} \qw & \ctrl{1} & \qw & \qw \rstick{$\ket{\psi_{z^2}}$} \\
		\lstick{$\ket{\psi_z}$} \qw & \targ{} & \meter{} & \cw \rstick{assert $0$}
	\end{quantikz}
	\caption{Implementing the arithmetic map $z \mapsto z^2$ in a quantum state's amplitudes.}
	\label{fig:complex-square}
\end{figure}

With a bit more work, we can also implement more complex arithmetic maps on $\overline{\mathbb{C}}$~\cite{kiss2006ComplChaosCondQubitDyn}. As an example, we show how to implement $z \mapsto z^2$. If we start with two copies of $\ket{\psi_z}$ and we apply a CNOT, then we obtain
\[\ket{\psi_z} \otimes \ket{\psi_z} \propto z^2\ket{00} + z\ket{10} + z\ket{01} + \ket{11} \xmapsto{\mathrm{CNOT}} z^2\ket{00} + \ket{10} + z\ket{01} + z\ket{11}.\]
If we now measure the second qubit in the computational basis and post-select on measurement outcome $0$, we obtain the state \[z^2\ket{0} + \ket{1} \propto \ket{\psi_{z^2}}.\]
This procedure is depicted in \cref{fig:complex-square}.

\subsubsection{Schr\"odinger's Microscope}

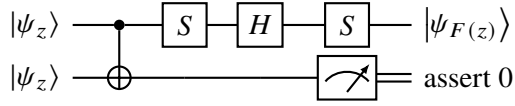
\begin{figure}[!t]
	\centering
	\begin{quantikz}[row sep={0.7cm,between origins},column sep=.45cm]
		\lstick{$\ket{\psi_z}$} \qw & \ctrl{1} & \gate{S} & \gate{H} & \gate{S} & \qw \rstick{$\ket{\psi_{F(z)}}$} \\
		\lstick{$\ket{\psi_z}$} \qw & \targ{} & \qw & \qw & \meter{} & \cw \rstick{assert $0$}
	\end{quantikz}
	\caption{The circuit that implements the Möbius transformation $F : z \mapsto \frac{z^2 + \ii}{\ii z^2 + 1}$.}
	\label{fig:F-circuit}
\end{figure}

Using these relations between complex analysis and quantum circuits, we wish to implement a non-trivial analytic transformation with a low-depth circuit.
To this end, we use the mapping $F : z \mapsto \frac{z^2 + i}{\ii z^2 + 1}$, which is implemented by the circuit~\cite{gilyen2015ExpSensInQPhys} displayed in \cref{fig:F-circuit}.

\begin{figure}[!t]
	\centering
	\begin{quantikz}[row sep={0.7cm,between origins},column sep=.45cm]
		\lstick{$\ket{\psi_z}$} \qw & \ctrl{1} & \gate{S} & \gate{H} & \gate{S} & \qw & \ctrl{2} & \gate{S} & \gate{H} & \gate{S} & \qw \rstick{$\ket{\psi_{F(F(z))}} = \ket{\psi_{F^{\circ 2}(z)}}$} \\
		\lstick{$\ket{\psi_z}$} \qw & \targ{} & \qw & \qw & \meter{} & \cw & \cw & \cw & \cw & \cw & \cw \rstick{assert $0$} \\
		\lstick{$\ket{\psi_z}$} \qw & \ctrl{1} & \gate{S} & \gate{H} & \gate{S} & \qw & \targ{} & \qw & \qw & \meter{} & \cw \rstick{assert $0$} \\
		\lstick{$\ket{\psi_z}$} \qw & \targ{} & \qw & \qw & \meter{} & \cw & \cw & \cw & \cw & \cw & \cw \rstick{assert $0$} \\
	\end{quantikz}
	\caption{The circuit that implements the $2$\textsuperscript{nd} level of Schr\"odinger's Microscope. The input $z \in \overline{\mathbb{C}}$ can be any point on the Riemann sphere. The generalization from level $n$ to $n+1$ is done by doubling the number of qubits and adding runs of the circuit in \cref{fig:F-circuit} accordingly. Thus, in level $n$, we use $2^n$ qubits and we run the circuit in \cref{fig:F-circuit} a total of $2^n-1$ times. Note that all the measurements can also be performed at the very end of the circuit.}
	\label{fig:SM-circuit}
\end{figure}
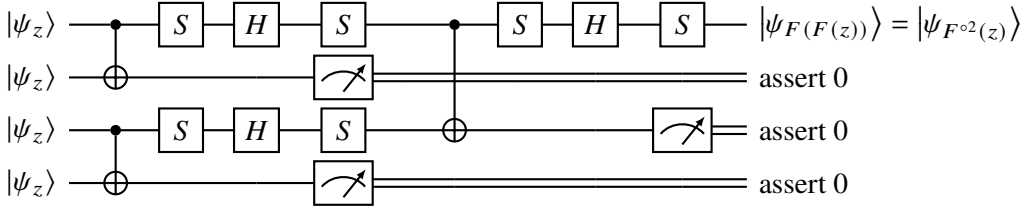

Next, we realize that it is also possible to implement $n$ consecutive applications of $F$, $F^{\circ n} = F \circ \cdots \circ F$, by multiplexing the outcomes of multiple circuits that implement $F$. This is displayed in \cref{fig:SM-circuit}. We call the $n$-times repeated transformation $F^{\circ n}$ the $n$\textsuperscript{th} level of Schr\"odinger's Microscope.

\begin{figure}[!t]
	\centering
	\begin{tabular}{cc}
		\includegraphics[width=.45\textwidth,draft=\hidepics]{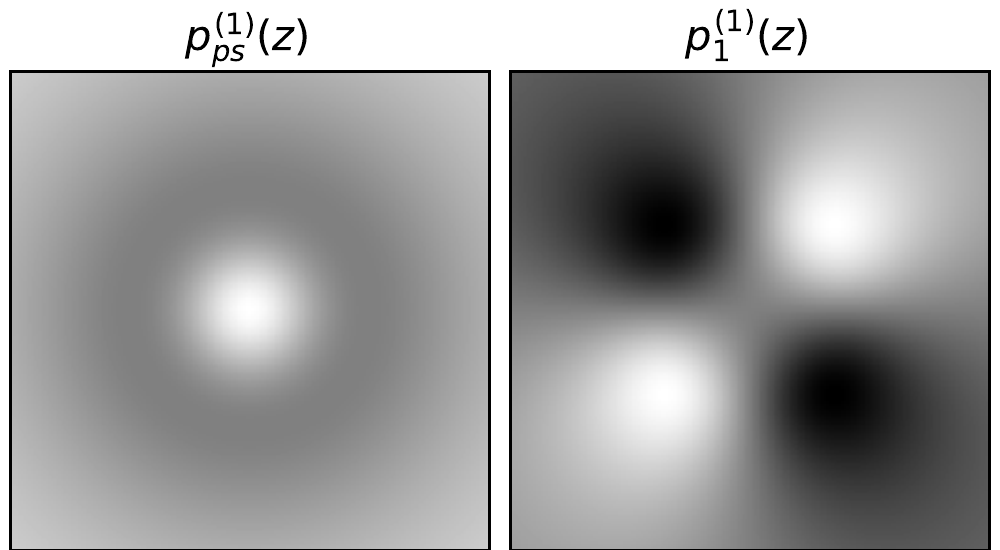} & \includegraphics[width=.45\textwidth,draft=\hidepics]{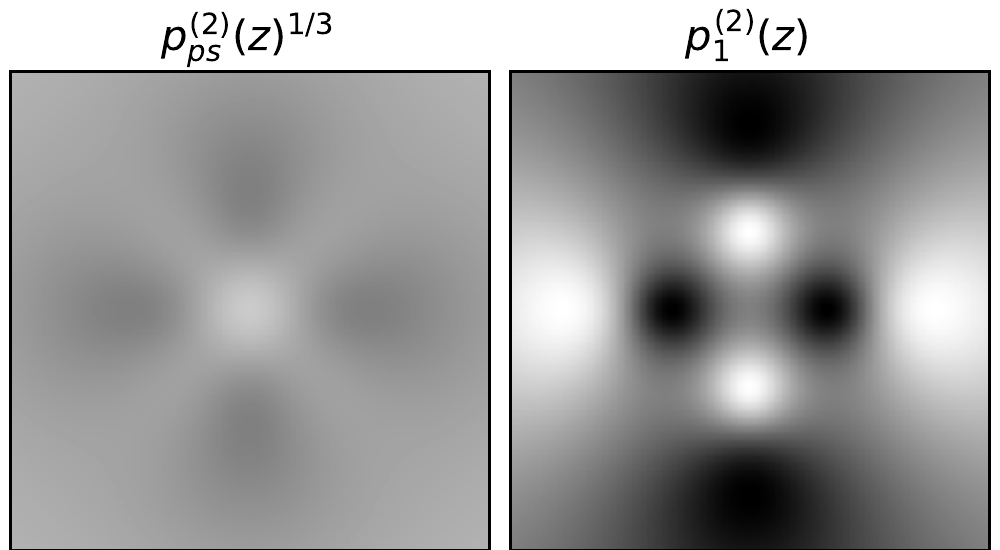} \\
		$n = 1$ & $n = 2$ \\
		\includegraphics[width=.45\textwidth,draft=\hidepics]{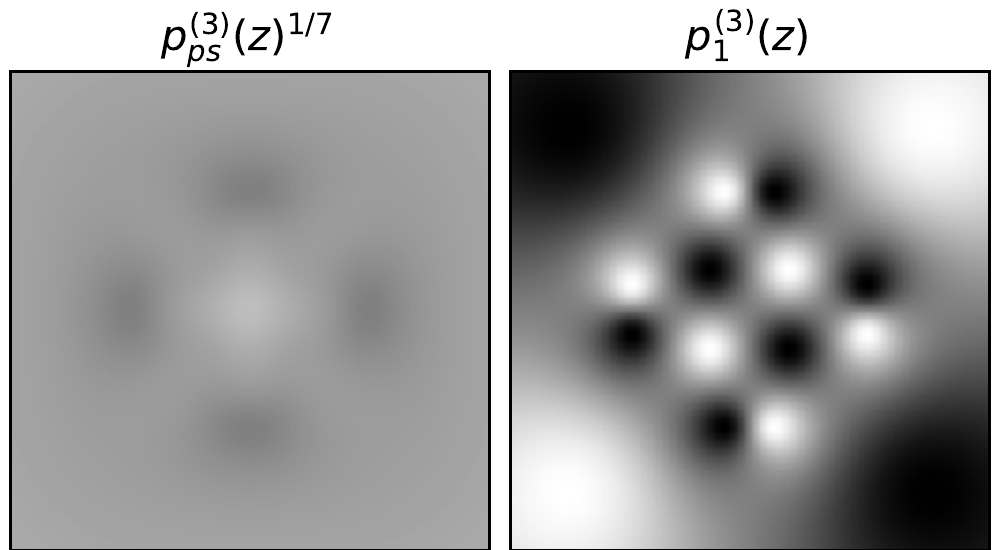} & \includegraphics[width=.45\textwidth,draft=\hidepics]{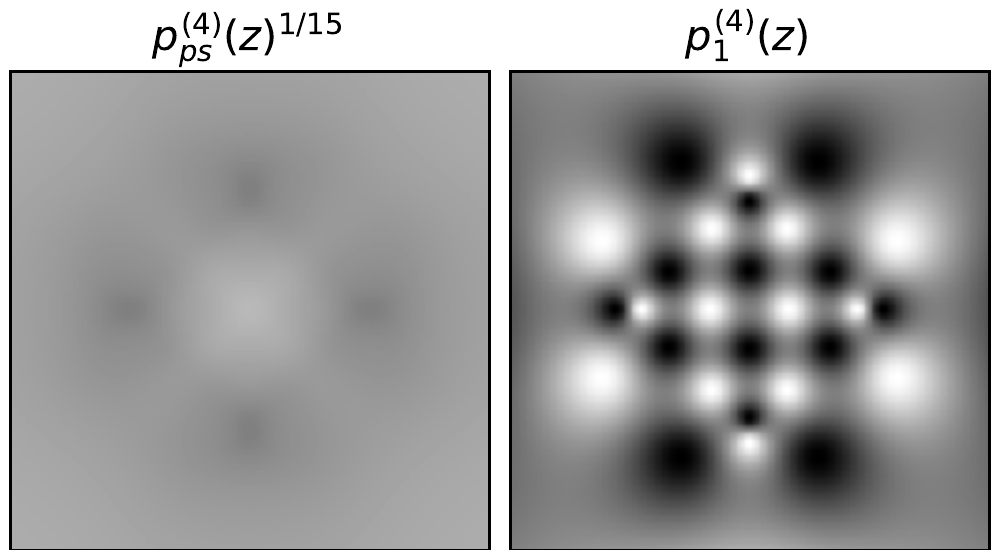} \\
		$n = 3$ & $n = 4$
	\end{tabular}
	\caption{The probability distributions associated to the $n$\textsuperscript{th} level Schr\"odinger's Microscope circuit shown in \cref{fig:SM-circuit}. Every pixel corresponds to a particular value for the complex input $z=x + \ii y$, with the real and imaginary parts $x$ and $y$ ranging from $-2$ to $2$ on the horizontal and vertical axes, respectively. We show the probabilities for $n = 1,2,3,4$. As for a given value of $n$ we perform $2^n-1$ steps of post-selection, we plot the geometric mean $p_{\text{ps}}^{(n)}(z)^{1/(2^n-1)}$ of all post-selection probabilities.}
	\label{fig:SM-simulation}
\end{figure}

We can associate two probabilities to the $n$\textsuperscript{th} level of Schr\"odinger's Microscope. First, we can calculate with what probability the overall post-selection succeeds, $p^{(n)}_{\text{ps}}(z)$, and second the probability that the first qubit is measured to be in the state $\ket{1}$, conditioned on the post-selection succeeding, $p_1^{(n)}(z)$. Those are given by
\[p_{\text{ps}}^{(n)}(z) = \prod_{j=1}^n \left(\frac{|F^{\circ(n-j)}(z)|^4 + 1}{(|F^{\circ(n-j)}(z)|^2 + 1)^2}\right)^{2^{j-1}} \qquad \text{and} \qquad p^{(n)}_1(z) = \frac{1}{|F^{\circ n}(z)|^2 + 1}.\]
For various levels $n$, we plot these probabilities depending on $z$ in \cref{fig:SM-simulation}.

The objective of the benchmark is to recreate the plots from \cref{fig:SM-simulation} using a quantum device. To that end, given a level $n$, we run the circuit in \cref{fig:SM-circuit} a fixed number of times, and approximate the probabilities using the resulting measurement frequencies. We quantitatively measure the quality of the resulting figure by calculating the root mean square difference with the analytically calculated picture displayed in \cref{fig:SM-simulation}. We refer to the resulting value as the \textit{score}, and hence the lower the score, the better the performance of the device.

As the probability plots in \cref{fig:SM-simulation} behave more and more oscillatory for larger and larger $n$, the more influence any noise in the quantum device will have on the outcome of the figure. Hence, we expect that at some $n$ the noise will start to dominate the figure outcome, and the higher the value of this $n$, the less noisy the device.

\subsubsection{Mandelbrot}

Similarly as in the Schr\"odinger's Microscope benchmark, we use the correspondence between the Riemann sphere and quantum states in this benchmark. This time, though, we construct a circuit~\cite{gilyen2015ExpSensInQPhys} in such a way that the probability plot resembles the Mandelbrot set.

To that end, recall that the Mandelbrot set is the set of values $c \in \mathbb{C}$ for which
\[\limsup_{n \to \infty} |G_c^{\circ n}(0)| < \infty, \qquad \text{where} \qquad G_c : z \mapsto z^2 + c.\]
Implementing the map $G_c$ requires a bit more work than $z \mapsto z^2$. A possible decomposition into elementary quantum gates is given in \cref{fig:Gc-circuit}.

\begin{figure}[!t]
	\centering
	\[\begin{quantikz}[row sep={0.7cm,between origins},column sep=.45cm]
		\lstick{$\ket{\psi_z}$} \qw & \ctrl{1} & \gate{H} & \ctrl{1} & \gate{Z^{\phi}} & \qw & \gate{R_2} & \ctrl{1} & \qw & \qw & \qw \rstick{$\ket{\psi_{z^2+c}}$} \\
		\lstick{$\ket{\psi_z}$} \qw & \targ{} & \ctrl{-1} & \gate{R_1} & \gate{Z^{-\phi}} & \gate{X} & \ctrl{-1} & \targ{} & \gate{X} & \meter{} & \cw \rstick{assert 0}
	\end{quantikz}\]
	where
	\[R_j = \frac{1}{\sqrt{1+r_j^2}}\begin{bmatrix}
		1 & r_j \\
		r_j & -1
	\end{bmatrix}, \qquad \frac{1}{r_1} = r_2 = |c| \cdot \sqrt{\frac12\left(1 + \sqrt{1 + \frac{4}{|c|^2}}\right)}, \qquad \phi = \frac{\arg(c)}{\pi}\]
	\caption{An implementation of $G_c$. Note that $R_j$ is a power of the $Y$-gate up to a global phase. Here $z \in \overline{\mathbb{C}}$ and $c$ can be any non-zero complex number. In the limit where $c \to 0$ we observe that the circuit reduces to a single CNOT.}
	\label{fig:Gc-circuit}
\end{figure}

\begin{figure}[!t]
	\centering
	\begin{tabular}{cc}
		\includegraphics[width=.45\textwidth,draft=\hidepics]{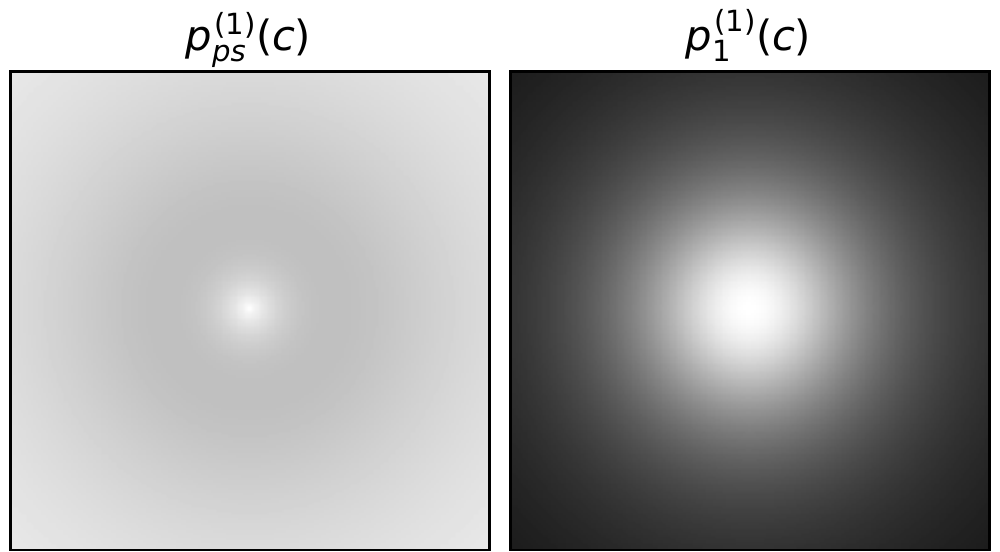} & \includegraphics[width=.45\textwidth,draft=\hidepics]{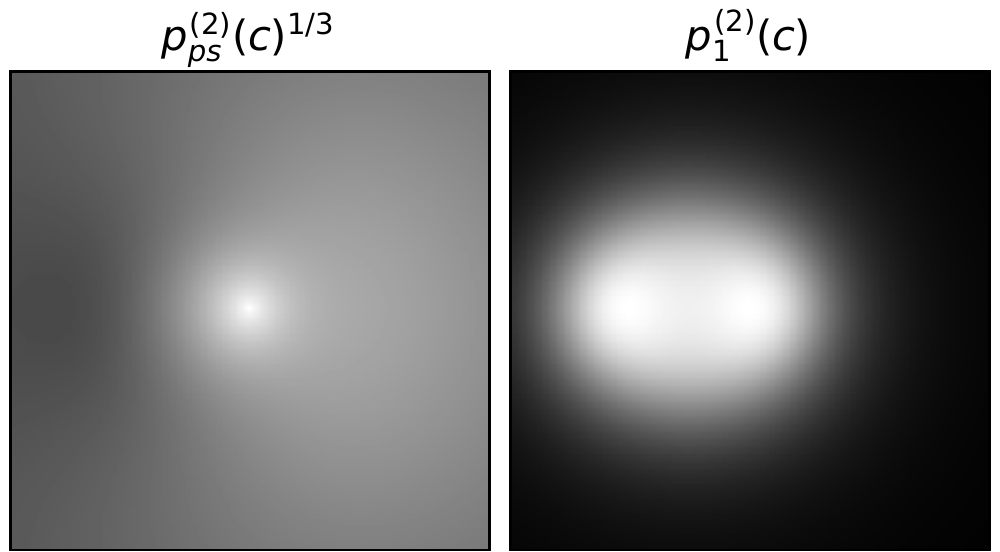} \\
		$n = 1$ & $n = 2$ \\
		\includegraphics[width=.45\textwidth,draft=\hidepics]{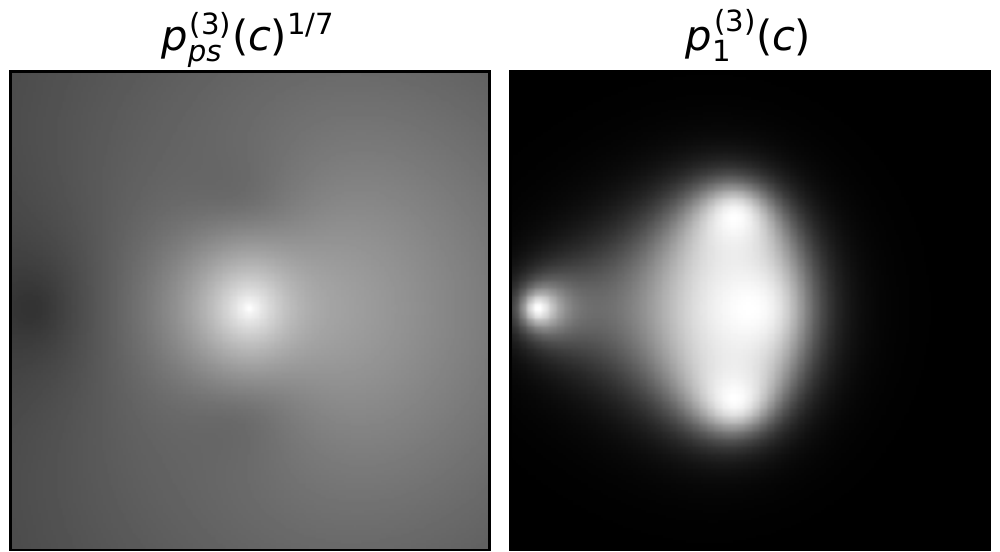} & \includegraphics[width=.45\textwidth,draft=\hidepics]{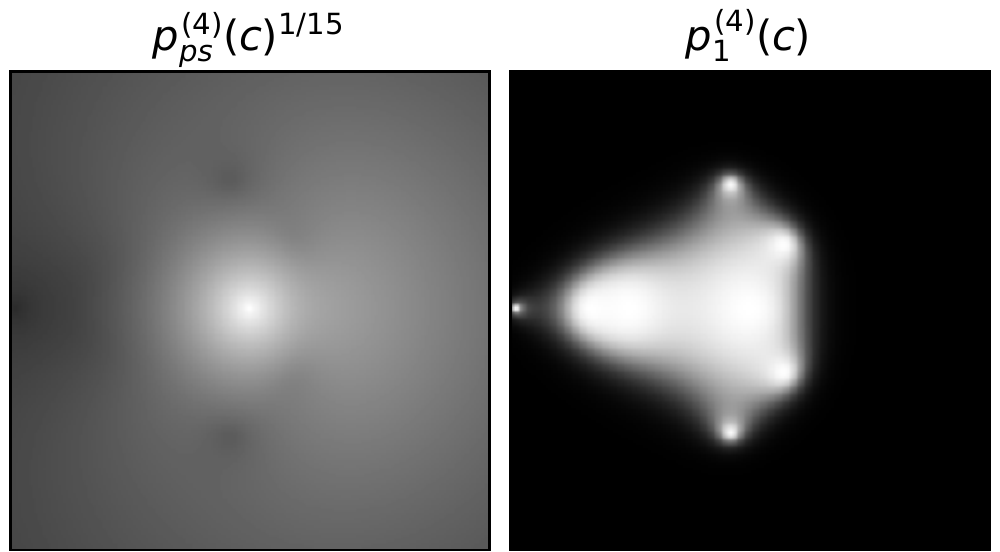} \\
		$n = 3$ & $n = 4$ \\
		\includegraphics[width=.45\textwidth,draft=\hidepics]{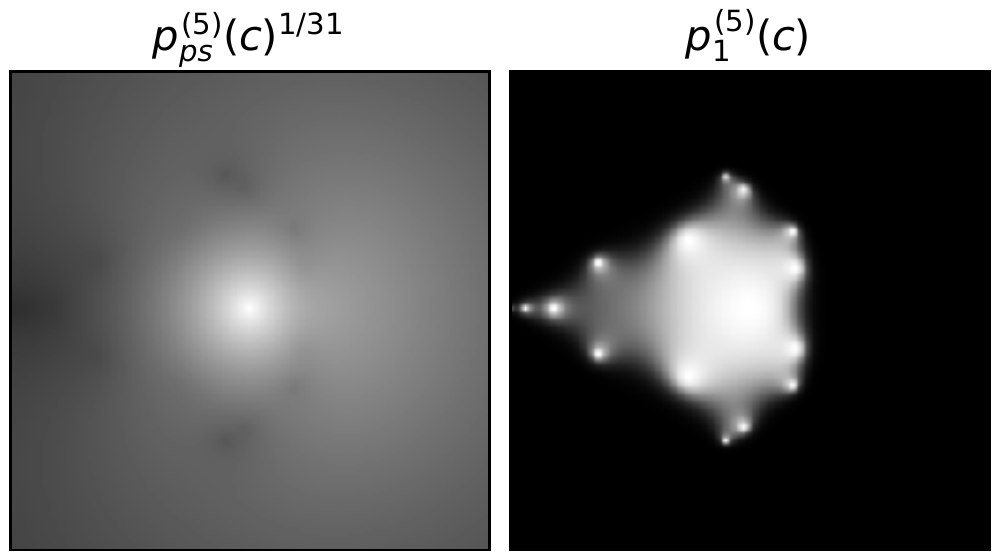} & \includegraphics[width=.45\textwidth,draft=\hidepics]{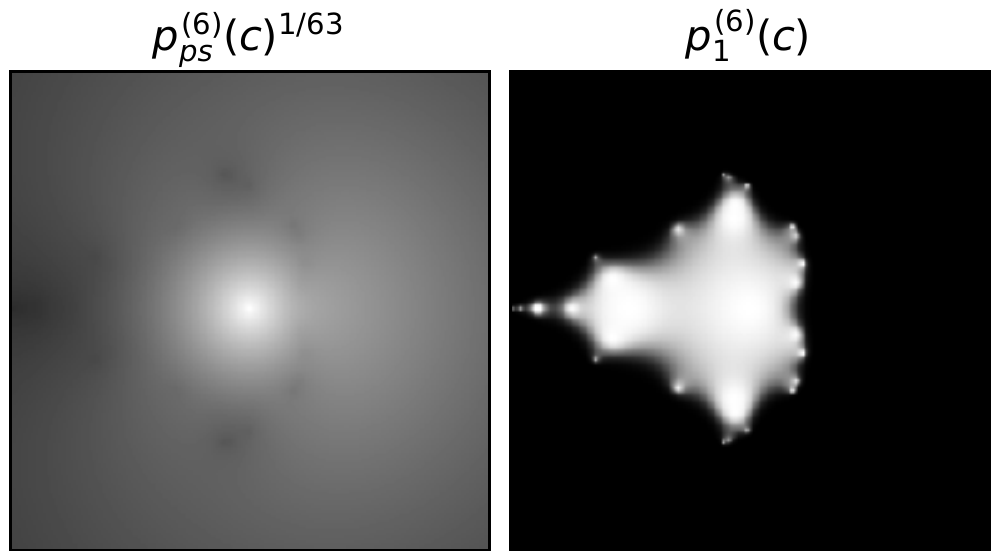} \\
		$n = 5$ & $n = 6$ \\
		\includegraphics[width=.45\textwidth,draft=\hidepics]{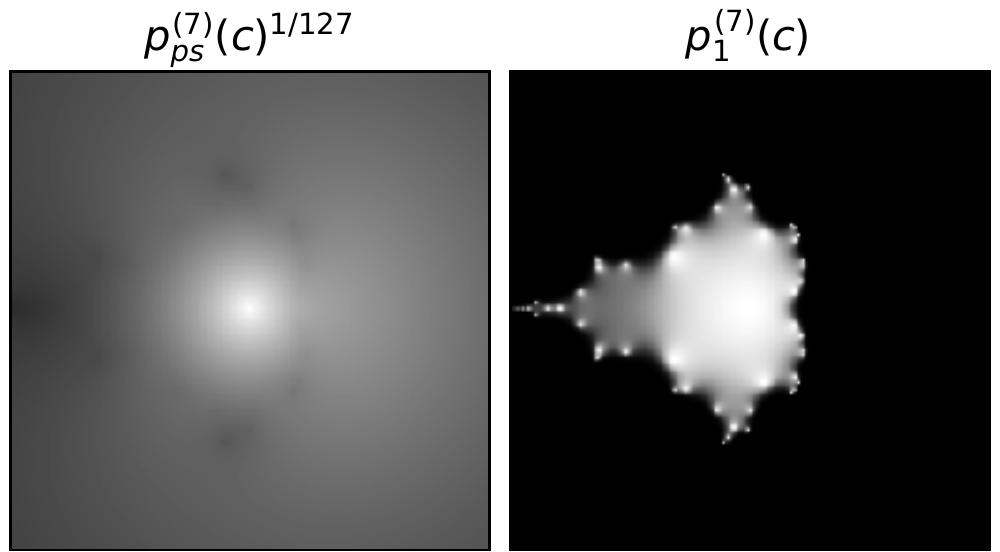} & \includegraphics[width=.45\textwidth,draft=\hidepics]{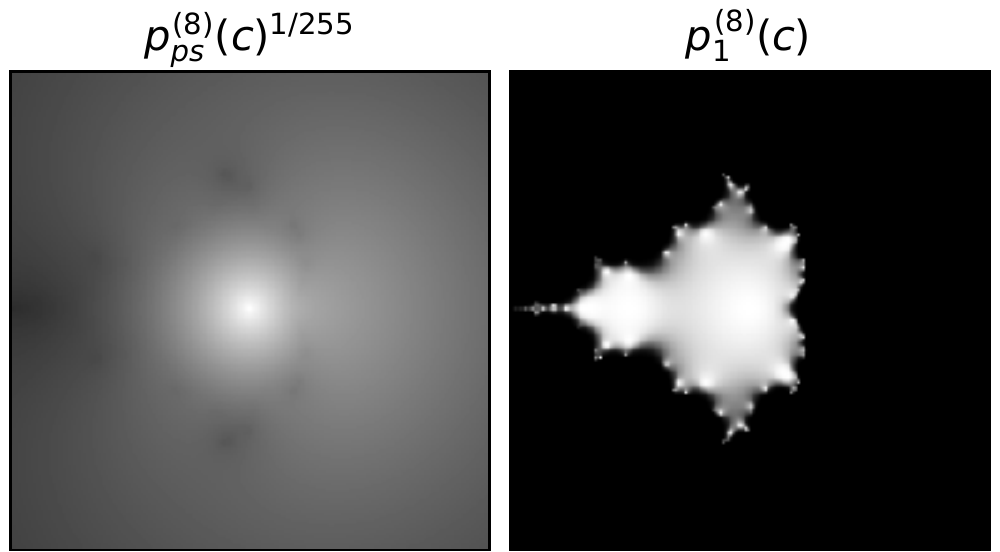} \\
		$n = 7$ & $n = 8$
	\end{tabular}
	\caption{The probability distributions associated to the $n$\textsuperscript{th} level Mandelbrot circuit. While in \cref{fig:SM-simulation} every pixel corresponds to a particular input state $z$, here every pixel corresponds to a particular value $c$ of the parameter in the complex map, with the real and imaginary parts ranging from $-2$ to $2$ on the horizontal and vertical axes, respectively. We show the probabilities for $n$ ranging from $1$ to $8$.}
	\label{fig:mandelbrot-simulation}
\end{figure}

Analogously to the $n$\textsuperscript{th} level Schr\"odinger's Microscope, we can define the probabilities $p_{\text{ps}}^{(n)}(c)$ and $p_1^{(n)}(c)$ for the $n$\textsuperscript{th} level of the Mandelbrot circuit.
This time, the probabilities depend on $c$ and we always use $\ket{\psi_0}^{\otimes 2^n} = \ket{1}^{\otimes 2^n}$ as the initial state.
We remark that
\[p_1^{(n)}(c) = \frac{1}{|G_c^{\circ n}(0)|^2 + 1},\]
and hence $c$ is an element of the Mandelbrot set if and only if
$\liminf_{n \to \infty} p_1^{(n)}(c) > 0.$
As expected we can see the Mandelbrot set appear in the plots of $p_1^{(n)}$ for larger values of $n$ in \cref{fig:mandelbrot-simulation}.

Similarly as before, the objective of the Mandelbrot benchmark is to reproduce the plots displayed in \cref{fig:mandelbrot-simulation} using measurement statistics obtained from running the circuit on a quantum device. We measure the quality of said device by calculating the root mean square difference between the resulting pictures and the analytically calculated ones from \cref{fig:mandelbrot-simulation}.

Many of the primary details of the Mandelbrot set become apparent only when $n$ is about $6$ or $7$, i.e., when the number of qubits used is between $64$ to $128$, with a corresponding circuit depth of $\approx 50$ to $100$. However, the post-selection probability decreases doubly-exponentially for increasing $n$, so in this regime one cannot hope to gather sufficient statistics to generate the right figure accurately. We remark that the number of runs required can be drastically reduced by running more instances in parallel and at each level collecting successful ``branches'' and/or employing amplitude amplification in between each round of post selection, at the cost of making the circuits more complex and deep. Since this is too complicated for current-day quantum devices we leave this modification of the circuits for future research, but we envision that it will be well-suited to test the properties of quantum devices in the late-NISQ era, where a significant number of qubits can be acted on with sufficiently high fidelity.

\subsection{Line Drawing}

The next benchmark that we propose is motivated by signal processing applications of quantum devices. Suppose that we have some $2^n$ points $z_0, \dots, z_{2^n-1} \in \mathbb{C}$ in the complex plane. If we draw these points in order and connect them by straight lines to form a closed loop, we draw a figure.
Equivalently, we can describe this figure by a set of $2^n$ Fourier coefficients, $c_0, \dots, c_{2^n-1} \in \mathbb{C}$, that define the sames set of points just in the Fourier domain, namely
\[z_t = \sum_{j=0}^{2^n-1} c_j\exp\left(2\pi \ii jt\right) \text{ for all } t \in \{0,1,\cdots,2^n-1\}.\]
In this benchmark, we use a quantum device to draw the points $z_0, \dots, z_{2^n-1}$ in the complex plane, given access to a quantum state whose state vector contains the Fourier coefficients as amplitudes. The deformation of the resulting figure tells us something about the noise that is present within the quantum device.

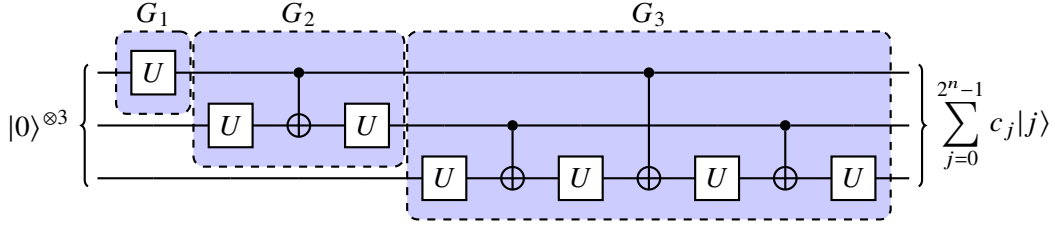
\begin{figure}[!t]
	\centering
	\begin{quantikz}[row sep={0.7cm,between origins},column sep=.45cm]
		\lstick[wires=3]{$\ket{0}^{\otimes 3}$} & \gate{U} \gategroup[1,steps=1,style={dashed,rounded corners,fill=blue!20, inner xsep=2pt},background]{$G_1$} & \qw \gategroup[2,steps=3,style={dashed,rounded corners,fill=blue!20, inner xsep=2pt},background]{$G_2$} & \ctrl{1} & \qw & \qw \gategroup[3,steps=7,style={dashed,rounded corners,fill=blue!20, inner xsep=2pt},background]{$G_3$} & \qw & \qw & \ctrl{2} & \qw & \qw & \qw & \qw\rstick[wires=3]{$\displaystyle \sum_{j=0}^{2^n-1} c_j\ket{j}$} \\
		& \qw & \gate{U} & \targ{} & \gate{U} & \qw & \ctrl{1} & \qw & \qw & \qw & \ctrl{1} & \qw & \qw \\
		& \qw & \qw & \qw & \qw & \gate{U} & \targ{} & \gate{U} & \targ{} & \gate{U} & \targ{} & \gate{U} & \qw 
	\end{quantikz}
	\caption{Structure of the state preparation circuit for $n = 3$, as described in \cite{BVMS04}. All the $U$'s denote distinct arbitrary single qubit gates. For arbitrary $n$, we generalize the circuit such that it has the building blocks $G_1, \dots, G_n$. Each block $G_j$ contains $2^j-1$ CNOT-gates.}
	\label{fig:state_preparation}
\end{figure}

The procedure consists of three steps. First, we supply the Fourier coefficients to our quantum device as the amplitudes in an $n$-qubit state, that we prepare through a state preparation circuit.
There has been quite some research in the direction of minimizing the CNOT-count of state preparation circuits, e.g.~\cite{BVMS04,SBM06,PB10}.
We use \cite{BVMS04}, which means that we prepare the $n$-qubit state containing the Fourier coefficients with $2^n-n-1$ CNOT gates, and $2^n$ single qubit unitaries. The circuit is displayed in \cref{fig:state_preparation}.

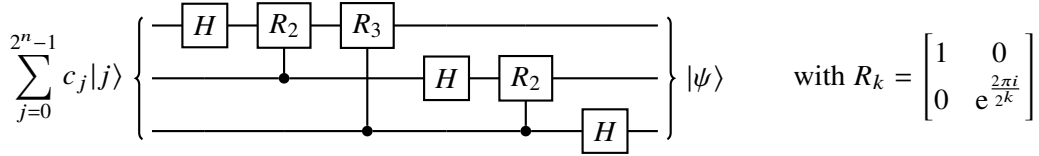
\begin{figure}[!t]
	\centering
	\vspace{-1em}
	\[\begin{quantikz}[row sep={0.7cm,between origins},column sep=.4cm]
		\lstick[wires=3]{\ensuremath{\displaystyle \sum_{j=0}^{2^n-1} c_j\ket{j}}}\qw & \gate{H} & \gate{R_2} & \gate{R_3} & \qw & \qw & \qw & \qw\rstick[wires=3]{\ensuremath{\ket{\psi}}} \\
		\qw & \qw & \ctrl{-1} & \qw & \gate{H} & \gate{R_2} & \qw & \qw \\
		\qw & \qw & \qw & \ctrl{-2} & \qw & \ctrl{-1} & \gate{H} & \qw 
	\end{quantikz} \qquad \raisebox{-1mm}{with $\displaystyle R_k = \begin{bmatrix}
		1 & 0 \\
		0 & \ee^{\frac{2\pi i}{2^k}}
	\end{bmatrix}$}\]
	\vspace{-2em}
	\caption{The circuit for the quantum Fourier transform for $n = 3$, following the standard construction, outlined for instance in \cite[Figure 5.1]{nielsen2002QCQI}. We drop the final SWAP-gates as we can simply interpret the order of the qubits as being reversed.}
	\label{fig:QFT}
\end{figure}

Next, we apply the quantum Fourier transform, which has a well-known circuit implementation, which can for instance be found in \cite[Figure 5.1]{nielsen2002QCQI}. In an attempt to reduce the number of gates, we do not include the final SWAP's in the circuit, but rather we reverse the order of the qubits by a classical post-processing step.
Neglecting normalization, this maps the state to
\[QFT_{2^n} \sum_{j=0}^{2^n-1} c_j\ket{j} = \sum_{j=0}^{2^{n-1}} z_j\ket{j} =: \ket{\psi},\]
which hence prepares the complex numbers that we want to draw as the  amplitudes in our state. The total number of two-qubit gates used in this step is $n(n-1)/2$, and the number of single qubit gates is $n$. For completeness, we include the circuit in \cref{fig:QFT}.

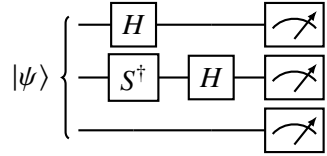
\begin{figure}[!t]
	\centering
	\vspace{-1em}
	\[\begin{quantikz}[row sep={0.7cm,between origins},column sep=.4cm]
		\lstick[wires=3]{\ensuremath{\ket{\psi}}}\qw & \gate{H} & \qw & \meter{} \\
		\qw & \gate{S^{\dagger}} & \gate{H} & \meter{} \\
		\qw & \qw & \qw & \meter{} \\
	\end{quantikz}\]
	\vspace{-2em}
	\caption{A circuit that measures in the eigenbasis of the Pauli string $X \otimes Y \otimes Z$. In general, we add $H$ and $HS^{\dagger}$ to all wires that correspond to $X$'s and $Y$'s in the Pauli string, respectively. If we start in an eigenstate of the Pauli string, i.e., if $\ket{\psi}$ is an eigenvector of the Pauli string, then in principle we can recover this eigenstate from the measurement outcome with $100\%$ probability.}
	\label{fig:pauli_measurement}
\end{figure}

Finally, in order to recover the points $z_j$, we estimate the amplitudes in the state vector, i.e., we perform full-state tomography under the assumption that the objective state is pure. Low-rank sate tomography is a well-established field, and we follow the approach outlined in \cite{GKKT18}. Every time we run the circuit, we perform a measurement in the eigenbasis of a randomly\footnote{For technical reasons in our implementations we measure each Pauli strings the same number of times in total.} selected Pauli string $P_1 \otimes \cdots \otimes P_n$, where $P_i \in \{X,Y,Z\}$. Performing such a measurement is achieved by appending at most $2n$ single qubit gates to the end of the circuit, and subsequently performing a standard basis measurement. This process is depicted in \cref{fig:pauli_measurement}.

In the $i$\textsuperscript{th} run, we denote the eigenstate corresponding to the measurement outcome of the Pauli measurement on the $j$\textsuperscript{th} qubit to be $\ket{\psi_i^{(j)}}$. After $N$ runs, we find that
\[L = \frac{1}{N}\sum_{i=1}^N \bigotimes_{j=1}^n \left(3\ket{\psi_i^{(j)}}\bra{\psi_i^{(j)}} - I\right)\]
is an unbiased estimator of $\ketbra{\psi}$, as proven in \cite[eq.~25]{GKKT18}. We estimate $\ket{\psi}$ by taking an eigenvector corresponding to the largest eigenvalue of $L$. Its coefficients, denoted by $\tilde{z}_j$, form the approximations to $z_j$'s.

Note that these estimates of $z_j$'s can only be accurate up to a global phase. Visually, this means that the resulting image will be rotated by an arbitrary angle. In order to be able to inspect how much the obtained image differ from the original, we rotate the image such that the $\ell_2$-distance between $\widetilde{z}$ and $z$ is minimized.
The remaining $\ell_2$-distance is the score of this benchmark weighted by the purity of the observed density operator, which implies that the lower the score, the better the quality of the device.

\begin{figure}[ht]
	\centering
	\includegraphics[width=\textwidth,draft=\hidepics]{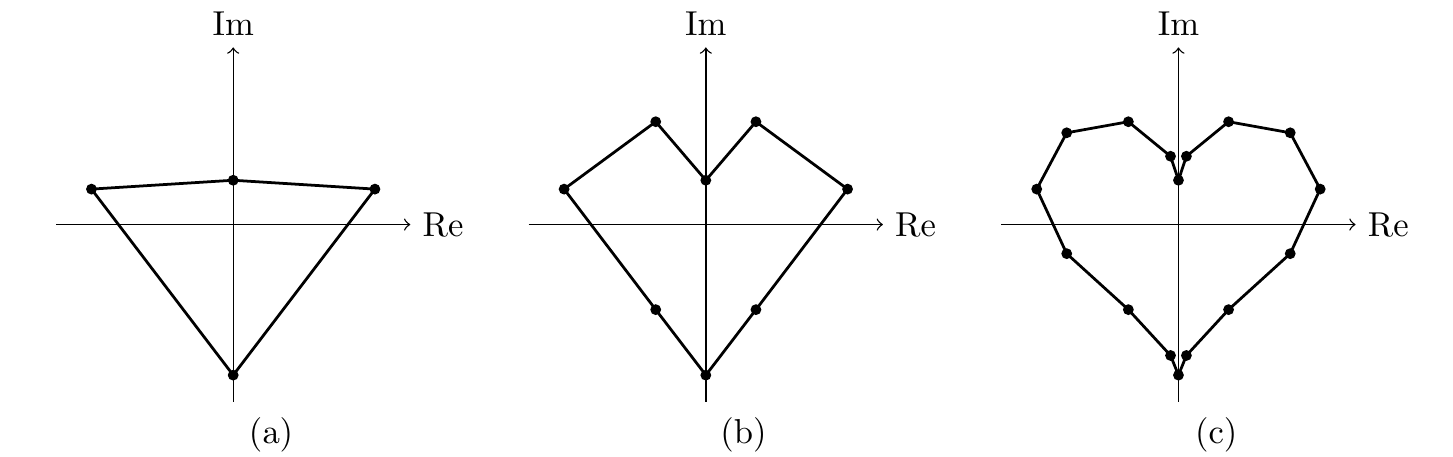}
	\caption{The curves that are used for testing the quantum devices. Curve (a) is referred to as the \emph{kite} can be drawn with a two-qubit quantum device. (b) and (c) can be drawn with three- and four-qubit quantum devices, respectively, and they are referred to as \emph{heart 8} and \emph{heart 16}. The points are sampled with evenly-spaced values for $t \in [0,2\pi]$ on the Lissajous curve defined by $x(t) = 16\sin(t)^3$, and $y(t) = 13\cos(t) - 5\cos(2t) - 2\cos(3t) - \cos(4t)$.}
	\label{fig:hearts}
\end{figure}

This method can be used on any closed curve defined by the set of points $z_0, \dots, z_{2^n-1} \in \mathbb{C}$. We define three specific closed curves that we use to benchmark current-day quantum devices. These are depicted in \cref{fig:hearts}.

\subsection{Quantum Matrix Inversion}
This benchmark is motivated by quantum signal processing \cite{low2016CompositeQuantGates,low2017HamSimUnifAmp} and quantum singular value transformation (QSVT) \cite{gilyen2018QSingValTransf} that are fundamental linear algebra primitives that capture the essence of many important quantum algorithms \cite{gilyen2018QSingValTransf}. QSVT takes as input a quantum circuit, and works with the bottom-right corner $A$ of the associated unitary matrix $U$---we call $U$ a \emph{block-encoding} of $A$:
\begin{equation*}
U= \left[\begin{array}{cc} . & . \\ . & A    \end{array}\right] \Longleftrightarrow A= \left(\bra{1}^{\otimes a}\otimes I\right) U \left(\ket{1}^{\otimes a}\otimes I\right).
\end{equation*} 
QSVT produces a new quantum circuit with associated unitary $U'$ where the singular values of the matrix $A$ are transformed by some bounded polynomial $P$, i.e., $U'$ is a block-encoding of $P^{(SV)}(A)$, where the $(SV)$ superscript is a reminder that polynomial is applied to the singular values rather than the eigenvalues of $A$.

\begin{figure}[!t]
	\centering
	\vspace{-1em}
	\[\begin{quantikz}[row sep={0.75cm,between origins},column sep=.25cm]
	& \gate{H}
	& \qw
	& \targ{}
	& \gate{\!\mathrm e^{\mathrm i\phi_{1} \sigma_z}\!}
	& \targ{}					
	& \qw	
	& \targ{}	
	& \gate{\!\mathrm e^{\mathrm i\phi_2 \sigma_z}\!}	
	& \targ{}			
	& \qw & \,\cdots\, &
	& \qw		
	& \targ{}
	& \gate{\!\mathrm e^{\mathrm i\phi_{d} \sigma_z}\!}
	& \targ{}		
	& \gate{H} 
	& \qw \\
	& \qw	
	& \gate[2]{\!U^{\dagger}\!\!}
	& \ctrl{-1}	
	& \qw
	& \ctrl{-1}				
	& \gate[2]{U}
	& \ctrl{-1}	
	& \qw
	& \ctrl{-1}
	& \qw & \,\cdots\, &
	& \gate[2]{\!U^{\dagger}\!\!}			
	& \ctrl{-1}	
	& \qw
	& \ctrl{-1}	
	& \qw
	& \qw  \\
	& \qwbundle[alternate]{}	
	& \qwbundle[alternate]{}
	& \qwbundle[alternate]{}
	& \qwbundle[alternate]{}
	& \qwbundle[alternate]{}	
	& \qwbundle[alternate]{}				
	& \qwbundle[alternate]{}
	& \qwbundle[alternate]{}
	& \qwbundle[alternate]{}
	& \qwbundle[alternate]{} & \,\cdots\, &
	& \qwbundle[alternate]{}		
	& \qwbundle[alternate]{}
	& \qwbundle[alternate]{}
	& \qwbundle[alternate]{}
	& \qwbundle[alternate]{}
	& \qwbundle[alternate]{}\\
	\end{quantikz}\]
	\vspace{-2em}
	\caption{The quantum circuit used for QSVT. The phases $\phi_j$ are determined by the polynomial $P$ to be applied to the singular values \cite{gilyen2018QSingValTransf}.}
	\label{fig:QSVT}
\end{figure}
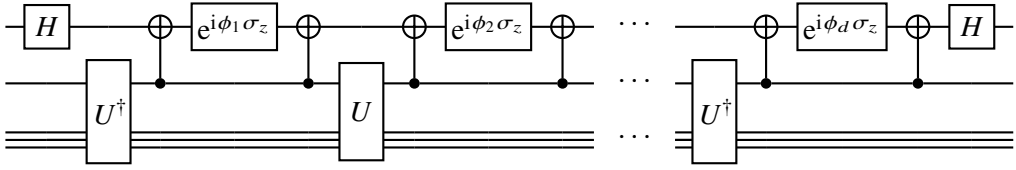

In particular the well-known HHL algorithm \cite{harrow2009QLinSysSolver} and its improved variants~\cite{ambainis2010VTAA,childs2015QLinSysExpPrec,chakraborty2018BlockMatrixPowers} 
can be cast as singular value transformation problem by using a polynomial that approximates the function $1/x$, 
enabling the construction of much more efficient quantum circuits compared to the notoriously difficult circuits used in the original paper \cite{harrow2009QLinSysSolver}. 
Indeed, in the easiest case when $a=1$, i.e., the matrix $A$ has exactly half the dimensions of $U$, and if we have a degree-$d$ polynomial approximation of $1/x$, then $U'$ consists of only $d$ applications of $U$ or $U^\dagger$, $2d$ CNOT gates, $d$ $Z$-rotations and $2$ Hadamard gates, see \cref{fig:QSVT}.

In order to make the used circuits as simple as possible we construct some short circuits $U$ encoding some non-trivial matrix $A$, with only two distinct singular values. Also we fine-tune the polynomial $P$ so that it only approximates $1/x$ on the actual singular values of $A$, in order to minimize its degree, and thereby the overall circuit depth. 

The procedure for a matrix $A$ of size $2^n \times 2^n$ is as follows. We take a block-encoding circuit $U$ of $A$, and apply singular value transformation to it, according to a polynomial that approximates $1/x$. Then, we apply the resulting circuit to the chosen computational basis state, and finally measure the state in the computational basis. This process requires the use of only $2$ auxiliary qubits (remember we set $a=1$). The outcome is only considered valid if the measurement on the first auxiliary qubit resulted in $0$ and $1$ on the second auxiliary qubit. In theory the measurement statistics of valid outcomes is proportional to the squared amplitudes of the columns of the inverted matrix $A^{-1}$. 

When we visualize the results, the $j$-th column depicts the histogram of valid outcomes corresponding to input state $\ket{j}$. Note that we do not employ post-selection, i.e., we do not normalize the histograms by the probability of getting valid measurement outcomes, therefore more noisy runs tend to produce more ``pale'' images. However, we do normalize colors so that the highest probability outcome in the ideal distribution would be $90$\% black, defining the color-scale. Since matrices of different size tend to produce different histograms, this means that the color-scale varies by the problem instance, and in the instances we present in particular $100$\% black corresponds to the following probabilities: $2\times 2: 0.592, 4\times 4: 0.868, 8\times 8: 0.130, 16\times 16: 0.210, 32\times 32: 0.316, 64\times 64: 0.261$. This explains why it is sometimes easier to achieve a less horrible score for a larger instance---the distribution may be farther from uniform for a smaller instance. Nevertheless, intuitively speaking achieving a really good score should be harder and harder for larger instances.

We quantify the quality of the resulting histograms by calculating the total variation distance between them and the ideal distributions corresponding to $A^{-1}$. The overall circuit depth is about $6n+20$, with approximately $4.5(n+2)(n+1)$ gates in total.

\subsection{Platonic Fractals}
It is well known that measurements can fundamentally change the state of a quantum system, and can lead to interesting non-linear dynamics when combined with post-selection~\cite{kiss2006ComplChaosCondQubitDyn,kiss2011MeasIndChaosEntangledSys,gilyen2015ExpSensInQPhys}, as we have seen in \cref{sec:Riemann}. 
Even without post-selection the dynamics can be rather interesting if one applies weak (or gentle) measurements. Unlike projective measurements, which project down the state to certain subspaces, weak measurements have a more nuanced effect. These measurements usually only extract some partial information from the system, and thereby disturb it less; as a byproduct repeated measurements can have different outcomes. 

Weak measurements are useful tools in quantum information theory, e.g., they can be used iteratively in order to prepare non-trivial ground states of certain local Hamiltonians~\cite{gilyen2016PrepGapHamEffQLLL}. Weak measurements can be implemented by using some entangling interaction with an ancillary system, followed by a projective measurement on the ancillas. Iteratively applying such operations give an interesting benchmark of the (strong) measurements implemented in the quantum system together with the employed entangling operation.

Certain weak measurements are such that they map pure states to pure states in the sense that, if the input is a pure state then the output is also some pure state depending on the measurement outcome ``label''. Thereby, if we iteratively apply the weak measurement and keep track of the observed measurement labels, then we can track down the trajectory of the pure state. As Jadczyk and Öberg~\cite{Jadczyk2002} pointed out, one can view such an iterative measurement process as a dynamical system on the set of pure states, which is the surface of the Bloch sphere in case the quantum system is a single qubit. Moreover, they showed that certain ``symmetric'' weak measurements, where the outcome labels correspond to vertices of a Platonic solid, lead to an associated fractal pattern on the surface of the Bloch sphere.

In the case of the octahedron the employed weak measurement is particularly easy to implement, so that it suffices to add a single ancilla qubit per iteration of the weak measurement. If one could reinitialize qubits on the fly, then this process could be indefinitely repeated by using only two qubits. However, currently available quantum hardware only allows for measurements at the end of the circuit, so the number of iterations is limited by the number of available qubits. Since in this benchmark we would like to track the pure state trajectories associated to the observed measurement labels, we apply some basic state tomography at the end of each run. 

\begin{figure}[!t]
	\centering
	\begin{subfigure}{.5\textwidth}
		\centering
		\[\begin{quantikz}[row sep={0.7cm,between origins},column sep=2mm]
		\lstick{\kern-1mm\ensuremath{\ket{0}}}\qw & \gate{\!\mathrm e^{\mathrm i \theta \sigma_x}\!} & \gate{\!\mathrm e^{\mathrm i(\frac\pi2-\theta) \sigma_x}\!} & \meter{} \\
		\lstick{\kern-1mm\ensuremath{\ket{\psi}}\!}\qw & \octrl{-1} & \ctrl{-1} & \qw
		\end{quantikz}\]
		\caption{Weak measurement circuit for angle $\theta$}		
		\label{subfig:weakcircuit}
	\end{subfigure}
	\begin{subfigure}{.5\textwidth}
		\centering
		$\left(
		\begin{array}{cccc}
		\cos (\theta ) & 0 & \mathrm i \sin (\theta ) & 0 \\
		0 & \sin (\theta ) & 0 & \mathrm i \cos (\theta ) \\
		\mathrm i \sin (\theta ) & 0 & \cos (\theta ) & 0 \\
		0 & \mathrm i \cos (\theta ) & 0 & \sin (\theta ) \\
		\end{array}
		\right)$
		\caption{The corresponding two-qubit unitary }	
		\label{subfig:weakuni}
	\end{subfigure}
	\caption{We use the above circuit for implementing our weak measurements in the Pauli $Z$-basis choosing $\theta\coloneqq\arccos\big(\sqrt{\frac{1+s}{2}}\big)$, where $s$ is the ``strength'' of the measurement. Indeed for $s=1$ we get a usual projective $Z$ (i.e., computational basis) measurement, while choosing $s=0$ does not give any information about the state, thereby leaving it undisturbed. The $X$ and $Y$ measurements are implemented similarly, by rotating the qubit state $\ket{\psi}$ to the corresponding basis.
	The outcome of iterated measurements is shown in \cref{fig:pf}.
	}
	\label{fig:weak}
\end{figure}

The particular octahedron measurement scheme for a strength $s$ and depth $d$ can be described as follows. First, prepare a single qubit in state $\ket{+}$. Choose $d$ Pauli bases (i.e., $X$, $Y$, or $Z$), and perform weak measurements in those bases consecutively, using the circuit described in \cref{fig:weak}. This process uses $d$ auxiliary qubits. Finally, (strongly) measure the original qubit in either the $Y$ or $Z$ basis. This scheme corresponds to the octahedron as the basis vectors of the three Pauli bases on the surface of the Bloch sphere correspond the vertices of the octahedron.

Using the techniques supplied in~\cite{Jadczyk2002}, we can describe what happens to the Bloch vector throughout this process. We start in state $\ket{+}$, hence the initial Bloch vector is $\vec{r}_0 = \vec{e}_x$. For each subsequent weak measurement in basis $b \in \{x,y,z\}$ with outcome $o \in \{0,1\}$, we can calculate the new Bloch vector via~\cite[Eq. (9)]{Jadczyk2002}:
\begin{equation}
	\label{eq:weak}
	\vec{r}_{j+1} = \frac{(1-k^2)\vec{r}_j + 2k(1+k(\vec{n} \cdot \vec{r}_j))\vec{n}}{1 + k^2 + 2k(\vec{n} \cdot \vec{r}_j)},
\end{equation}
where
\[k = \frac{1-\sqrt{1-s^2}}{s}, \qquad \text{and} \qquad \vec{n} = (-1)^o\vec{e}_b.\]
Since there are $3^d$ Pauli strings of length $d$, and we have $2^d$ possible outcomes per measurement string, every run of the above procedure yields one of $6^d$ possible outcomes. For each of these, we estimate the $Y$ and $Z$ components of the resulting Bloch vector, using the measurement statistics of the final measurement. The resulting plot contains the obtained estimated trajectories of these $6^d$ paths.

In order to quantify the quality of the resulting plot, we calculate the $\ell^2$-norm between the estimated points, and the analytically calculated Bloch vectors using \cref{eq:weak}. The circuits use at most $2+d(6\pm 2)$ gates and have circuit depth at most $2+5d$, where $d$ is the number of measurement directions used.

\begin{figure}[!t]
    \centering
    \includegraphics[width=\textwidth]{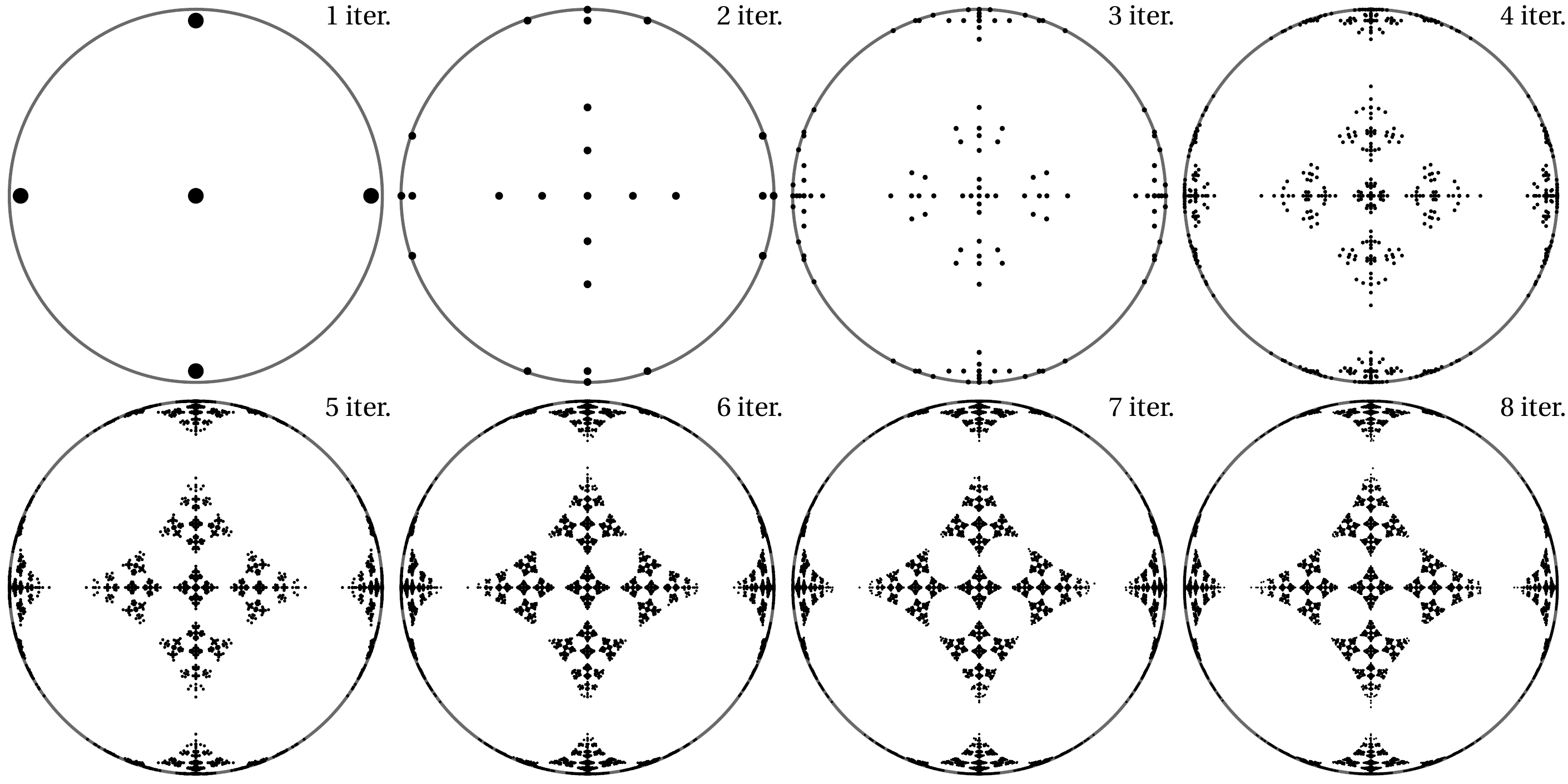}
    \caption{Platonic fractals for $n=1,\ldots,8$ iterations, and a weak measurement strength of $s=75\%$.
    In \cref{sec:results}, we show the points connected by paths which highlight their measurement sequence; in effect creating an $L$ system that branches sixfold at every node. This is helpful to visualize noisy benchmarking runs; however, as explained in \cref{sec:scalability}, for higher-fidelity devices we propose reverting to just drawing points, and collating them in a pixel grid (instead of running all exponentially many measurements), rendering the benchmark scalable.
    }
    \label{fig:pf}
\end{figure}

As shown in \cref{fig:pf}, for larger and larger iteration numbers, the points fill in a fractal on the Bloch sphere.

\subsection{Scalability of Benchmarks}\label{sec:scalability}
As our aim is to provide a suite of benchmarks that does not only work in the near term, but also in the mid- to long run (i.e.\ up to the point when fully error-corrected quantum hardware is available, or even further), we need to explain how our benchmarks scale to larger devices.
Efficiently scalable, in this setting, simply means that with larger devices the benchmarks can still be performed \emph{in a meaningful manner}, even though slight modifications might be necessary in how the circuits are run or the scores are obtained.

\paragraph{Bell Test.}
The Bell test scales quadratic in the number of qubits in the device and only requires a single basis measurement, as shown in \cref{tab:settings}; alternatively one can restrict the distance of qubits-to-be-tested, further reducing the number of required settings.\footnote{One could argue whether a quadratic runtime for an algorithm is truly ``scalable'', in the strictest sense of the word. Our take is that a device with 100 qubits should be able to take sufficient measurement statistics for $100^2$ circuits in a reasonable time; and a device with a ten thousand qubits should be able to do the same for a hundred million runs.}

\paragraph{Schrödinger's Microscope and Mandelbrot.}
In case of the Schrödinger's Microscope circuit each qubit measurement has post-selection probability $\Theta(1)$, therefore the overall post-selection probability is typically exponentially small in the number of qubits. However, this probability of one step turns out to be always at least $\frac{1}{2}$~\cite{gilyen2015ExpSensInQPhys}, and so out of $k$ pairs of qubits we get at least $k/2$ successful post-selections in expectation. Therefore, if we have about $4^{\ell}$ qubits at level $\ell$, then by collecting the successful outcomes at each level, we can ensure that the post-selection probability is large. It is not too difficult to collect the successful branches, for example using reversible sorting networks~\cite{gilyen2014MScThesis}, providing an efficient and low-depth implementation that no longer has the diminishing success probability issue. The resulting circuits can be efficiently constructed and executed, and the overall circuit depth stays poly-logarithmic in the number of qubits (assuming all-to all connectivity).\footnote{The number of qubits needed to perform $\ell$ iterations is still exponential, but for a given number of qubits (where the iteration number is necessarily logarithmic), the number of required operations is polynomial, making this a scalable benchmark.}
A similar approach can also be employed for the Mandelbrot benchmark, and optionally one could employ additional means of improving the post-selection probability, for example by using amplitude amplification at each level.

\paragraph{Line Drawing.}
The circuits in the line drawing benchmark on $n$ points contain $\Theta(n)$ gates and run on $\Theta(\log(n))$ qubits, and in order to obtain sufficient measurement statistics it suffices to run these circuits for a total number of shots that is polynomial in $n$. Thus, this benchmark scales well in the number of points that comprise the curve, and primarily tests the performance of quantum devices on deep and narrow circuits. In addition, this benchmark can be easily adapted to test more features of quantum devices. For instance, instead of preparing a quantum state that contains the Fourier coefficients of the curve as its amplitudes, one can also prepare the curve itself in the amplitudes of the quantum state, then recover the Fourier coefficients by measuring in the Fourier basis, and finally use classical post-processing by Fast Fourier Transform to reconstruct the curve from these. If additionally one has an analytical description of the curve to be drawn, one could substitute the state preparation circuit for a method that analytically calculates the points of the curve in the quantum state's amplitudes~\cite{grover2002SuperposEffIntegrProbDistr}. This would yield a somewhat more intricate circuit, and test the device's ability to perform computations in superposition, which in longer-term applications is an important element in quantum algorithm design. In addition to the number of points that are to be sampled, the difficulty of the resulting test can now be varied by controlling the intricacy of the analytical representation of the curve. 

Another variant of the benchmark that has a depth that scales polynomially with the number of qubits $k$, is to prepare a periodic quantum state with period $\Theta(k)$, which is then transformed using the quantum Fourier transform. The output state is then expected to have only $\Theta(k)$ non-zero amplitudes, so tomography becomes also scalable, and the resulting image is supposedly a cycle on $\Theta(k)$ points. This variant would test large Fourier transforms with relatively shallow circuits.

\paragraph{Quantum Matrix Inversion.}
The HHL algorithm is appealing because it can work with matrices that are exponentially large. This also means that it rapidly becomes infeasible to plot the entire matrix. Instead, we could plot a fixed size submatrix spanned by randomly chosen (basis) vectors. There are two other difficulties with scalability. Fist, the problem quickly becomes classically intractable, preventing the verification of the outcomes. Second, it is not immediately clear how to find shallow circuits that give rise to block-encodings with bounded condition numbers.

We can solve the condition number issue by taking a random shallow circuit and transforming its singular values by a polynomial with large constant term~\cite{gilyen2018QSingValTransf}. In order to get an easily verifiable outcome we could invert the block-encoding, and then multiply together the block-encoding with its (approximate) inverse---the resulting block-encoding is expected to be proportional to identity. The modified benchmark would thus just depict a randomly chosen submatrix of a block-encoding that is supposedly proportional to the identity, resolving all scalability issues regarding circuit design and verification.

\paragraph{Platonic Fratals.}
Currently, the platonic fractal benchmark is visualized as an L-system; lines connect the successive measurement spots within a 2D projection of the Riemann sphere.
As highlighted in \cref{tab:settings}, each additional iteration raises the number of possible paths by a factor of six; as such, the benchmark appears non-scalable.

If we instead cover the projected sphere as shown in \cref{fig:pf} by a pixel grid and collect all those outcomes that we expect to land within a single pixel (which we know analytically by the sequence of weak measurement outcomes), we can collate the collective measurement statistics of such outcomes to obtain the actual measured location of these points, which then might fall into the interior of the Bloch sphere.
If a bin contains less than some threshold number of measured paths (say, 100) so that we do not have sufficiently many measurement outcomes, then we discard it; otherwise we draw it at its measured position.
If the device is noisy, we expect a shrinked or blurry distortion (just like the images look like now). On the other hand, if the device is good, then the fractal should emerge on the surface of the Bloch sphere.
In this case, the scoring can be done with respect to the difference between the expected and measured coordinates for the bins that are not discarded.
The benchmark thus does no longer scale exponentially with the number of fractal iterations, but is governed by with the number of desired samples (as we do not require all possible paths to be measured) and the resolution of the image grid.

\pagebreak
\section{Testing IBM, Rigetti and IonQ Quantum Hardware}\label{sec:results}

Our benchmarking suite implements the \emph{same} tests for \emph{all} devices that we target: 
IBM (Qiskit / IBMQ), Rigetti (Forest / QCS and Amazon Braket), and IonQ (Amazon Braket), 
with the possibility of including other devices in the future; a simple command line program allows us to spawn the benchmarks, run the standardized tests, and visualize and score the outcomes in one go. In addition, we implemented the capability of simulating each benchmark exactly on a classical computer, either by computing the full state vector and calculating the resulting measurement statistics from it directly, or by emulating runs of the circuits on an ideal device, such that only statistical errors caused by having only finitely many measurement shots contribute to the imperfections in the resulting figure.

This unified interface allows us to establish a fair comparison, avoiding user-side optimizations that are vendor-specific; for each circuit, we leave it to the compiler to make an optimal choice with regards to choice of gate decomposition and qubit placement, and we do not manually interfere with this process.
The only exception is when we test a circuit with respect to a specific subset of qubits (such as a Bell test); in this case, we pin the endpoints, and choose the best routing between those endpoints based on the gate fidelities reported back from the device specifications. 

From 2019 till 2021, we tested 16 IBM machines, four Rigetti devices, and an IonQ device with an average of about 50 million individual circuit runs per machine (for an overall of approximately one billion experiments).

For each benchmark and device, we collect data with the settings presented in \cref{tab:settings}. In this section, we present the results of these experiments.
We discuss in \cref{subsec:scores} how we quantify the performance for each benchmark, and in \cref{sec:errors,sec:disentangling} how uncertainties are calculated, finally in \cref{subsec:results} we present the device-specific results.
In \cref{sec:meanscores} we evaluate mean scores for those devices where all six benchmarks were executed, and in \cref{sec:qv} we discuss to what extent our scores match the reported quantum volume.

\begin{table}[!tb]
	\hspace*{-0.9cm}
	\begin{tabular}{rllcllll S[table-format=3.2]}
		\toprule
		Benchmark & Settings & \makecell[l]{\# Individual\\\ \ circuits} & \makecell[l]{\# Meas.\\\ \ \ bases} & {\# Gates} & \makecell[l]{Circuit \\ depth} & \makecell[l]{\# Shots\\ \ \ \ per circuit} & {\# Runs ($\times10^6$)} \\
		\midrule
		\multirow{3}{*}{\makecell{Bell Test}}
		& & \multirow{3}{*}{$2\binom{\mathrm{\# qubits}}{2}$ pairs} & \multirow{3}{*}{3} & \multirow{3}{*}{$2\mathrm{dist}+1$} & \multirow{3}{*}{$2\mathrm{dist}+1$} & \multirow{3}{*}{8192} & 0.5 {\footnotesize{(5 qubits)}} \\
		& & & & & & & 3.8 {\footnotesize{(13 qubits)}} \\
		& & & & & & & 22.9 {\footnotesize{(31 qubits)}} \\[3mm]
		\multirow{3}{*}{\makecell[r]{Schrödinger's\\Microscope}}
		& $n=1$ & $32\times32$ pixels & 1 & $8$ & $6$ & 4096 & 4.2 \\
		& $n=2$ & $32\times32$ pixels & 1 & $20$ & $10$ & 4096 & 4.2 \\
		& $n=3$ & $32\times32$ pixels & 1 & $44$ & $14$ & 8192 & 8.4 \\[3mm]
		\multirow{3}{*}{Mandelbrot}
		& $n=1$ & $32\times32$ pixels & 1 & $13$ & $11$ & 4096 & 4.2 \\
		& $n=2$ & $32\times32$ pixels & 1 & $37$ & $21$ & 4096 & 4.2 \\
		& $n=3$ & $32\times32$ pixels & 1 & $85$ & $31$ & 8192 & 8.4 \\[3mm]
		\multirow{3}{*}{Line Drawing}
		& kite     & $25$ lines & $3^2$ & $6 \pm 1$ & $6 \pm 1$ & 4096 & 0.9 \\
		& heart 8  & $25$ lines & $3^3$ & $16 \pm 2$ & $12 \pm 1$ & 4096 & 2.8 \\
		& heart 16 & $25$ lines & $3^4$ & $35 \pm 3$ & $26 \pm 1$ & 4096 & 8.3 \\[3mm]
		\multirow{6}{*}{\makecell[r]{Quantum\\Matrix\\Inversion}}
		& $2 \times 2$ & $2$ columns & 1 & $26 \pm 1$ & 20 & 8192 & 0.02 \\
		& $4 \times 4$ & $4$ columns & 1 & $57 \pm 1$ & 32 & 8192 & 0.03 \\
		& $8 \times 8$ & $8$ columns & 1 & $90 \pm 2$ & 38 & 8192 & 0.07 \\
		& $16 \times 16$ & $16$ columns & 1 & $118 \pm 2$ & 44 & 8192 & 0.13 \\
		& $32 \times 32$ & $32$ columns & 1 & $152 \pm 3$ & 50 & 1024 & 0.03 \\
		& $64 \times 64$ & $64$ columns & 1 & $203 \pm 3$ & 56 & 1024 & 0.07 \\[3mm]
		\multirow{3}{*}{\makecell[r]{Platonic\\Fractals}}
		& 1 step  & $3$ directions  & 2 & $7 \pm 2$ & $5 \pm 2$ & $131072$ & 0.9  \\
		& 2 steps & $9$ directions  & 2 & $13 \pm 4$ & $8 \pm 4$ & $131072$ & 3.6  \\
		& 3 steps & $27$ directions & 2 & $19 \pm 6$ & $11 \pm 6$ & $131072$ & 10.6 \\
		\bottomrule
	\end{tabular}
	\caption{Experimental settings for all benchmarks.}
	\label{tab:settings}
\end{table}

\subsection{Scores}
\label{subsec:scores}

In the previous section, we have already explained how one can quantify the performance of a quantum device for each of the benchmarks. In this section, we explain the test-specific considerations for calculating these scores, and we also elaborate on the uncertainty analyses.

\paragraph{Bell Test.}
We collect measurement data needed for Bell coefficients $\CBell(A,B)\in[0,3/2]$ for any pair of qubits $A,B$ on the device.
As we expect the device interactions to feature a direction-dependent fidelity, we generally expect $\CBell(A,B) \neq \CBell(B,A)$.
For each device to test, we plot the matrix $\CBell(A,B)$ as a heatmap. 
$\CBell$ is a sum of Bernoulli random variables, for each of which we have that the maximum likelihood estimator is simply given by the mean of the $n$ counts $\mu = \sum_i x_i / n$, with variance $\Delta C^2 = \mu(1-\mu)/n$.
The standard deviation for the $\CBell$ values in the heatmap then follows from error propagation, i.e.
\[
\Delta\!\CBell = \sqrt{ \Delta C(0,2\pi/3)^2 + \Delta C(0,\pi/3)^2 + \Delta C(\pi/3,2\pi/3)^2 }.
\]

To each device, we also associate a total score, where we only count the nearest-neighbour Bell violation capabilities: we average all $\CBell$'s for the $N$ directly-connected pairs of qubits in the device, and take the resulting standard deviation, i.e.\ for neighbouring qubits $A\sim B$, we write
\[
    \mathrm{score} = \frac1N \sum_{A \sim B} \CBell(A,B) \pm \sqrt{ \frac1N \sum_{A\sim B} \Delta\!\CBell(A,B)^2 }.
\]
A perfect device would consistently produce a maximal Bell violation, and henceforth achieve a score of $3/2$.

\paragraph{Schrödinger's Microscope \& Mandelbrot.}

For a run of the $n$th level of the Schr\"odinger's Microscope benchmark, we run circuits on $2^n$ qubits that contain $2 \cdot 2^n + 4 \cdot (2^n-1)$ gates, and have depth $2 + 4n$. Similarly, for a run of the $n$th level Mandelbrot benchmark, we use $2^n$ qubits, $2^n + 11 \cdot (2^n-1)$ gates and depth $1 + 10n$. Since the number of gates grows much more rapidly than the circuit depth, these benchmarks test the performance of the device on shallow circuits.

The score associated to a single image is the mean-squared distance between itself and its analytically calculated counterparts. Each experiment yields two pictures, containing $p \times p$ pixel values ranging between $0$ and $1$, say $p_{i,j}$ and $q_{i,j}$, at locations $z_{i,j} \in \mathbb{C}$ in the complex plane. Hence, the resulting scores are calculated by
\begin{align*}
\text{score}_\mathrm{ps} &= \sqrt{\frac{1}{p^2} \sum_i\sum_j \left(p_{i,j} - p_\text{ps}(z_{i,j})^{\frac{1}{2^n-1}}\right)^2}, \\
\text{score}_1 &= \sqrt{\frac{1}{p^2} \sum_i \sum_j \left(q_{i,j} - p_1(z_{i,j})\right)^2}.
\end{align*}
In the post-selection probability plot, the value of $p_{i,j}^{2^n-1}$ is determined by the fraction of experiments where the measurement outcome on the post-selection qubits was the all zeros-string. Since all of these runs are independent, we obtain that the standard error for $p_{i,j}^{2^n-1}$ is estimated by the square root of the variance of a binomial distribution:
\[\Delta(p_{i,j}^{2^n-1}) = \sqrt{\frac{p_{i,j}^{2^n-1}(1-p_{i,j}^{2^n-1})}{\#\text{shots}}}.\]
We then calculate the standard error of $\text{score}_{\text{ps}}$ by standard error propagation.

Similarly, in the success probability plot, the value of $q_{i,j}$ is determined by the fraction of experiments where the measurement outcome of the first qubit is $1$, among all of the experiments where the post-selection succeeded. Since this latter is $p_{i,j}^{2^n-1} \cdot \#\text{shots}$, we find that the standard error of $q_{i,j}$ can be estimated by
\[\Delta q_{i,j} = \sqrt{\frac{q_{i,j}(1-q_{i,j})}{p_{i,j}^{2^n-1}\#\text{shots}}}.\]
The standard error of $\text{score}_1$ can now also be readily calculated from these estimates through standard error propagation.

Every device that is able to run the first two levels of the Schr\"odinger's Microscope and Mandelbrot benchmarks also receives an overall score. It can be calculated by simply taking the average of all the scores for the $1$st and $2$nd level runs.

Since even a perfect device will suffer from statistical errors in the measurement outcomes, there is a limit to the scores that quantum devices can hope to achieve with a fixed number of runs. For comparison, we have also included a simulated run alongside with the actual results.

\paragraph{Line Drawing.}
A run of the Line Drawing benchmark with $2^n$ points requires $n$ qubits to run, and the resulting circuits contain between $4 \cdot 2^n - n - 3 + \frac12n(n+1)$ and $4 \cdot 2^n + n - 3 + \frac12n(n+1)$ gates. Their depth is between $4 \cdot 2^n - 2n - 1$ and $4 \cdot 2^n - 2n + 1$.

We score each batch of experimental results the following way: let $\ket{\tilde{\psi}}$ be the eigenvector (pure state) corresponding to the largest eigenvalue (probability) $p$ in the estimated density operator, then we define the score as $(1-\sqrt{p})+\sqrt{p}\min_{\phi}\lVert\ket{\psi}-\ee^{\ii\phi}\ket{\tilde{\psi}}\rVert$, where $\ket{\psi}$ is the pure quantum state corresponding to the original curve. We collect data in $25$ batches, and take their average. The standard error is inferred statistically from the scores obtained for each of these batches.

To every device that is able to run the $4$ and $8$ point version of this benchmark, we associate an overall score which is simply the average of the scores obtained with $4$ and $8$ points. The standard error follows by regular error propagation.

Even an ideal device would not recover the exact figure due to statistical effects induced by performing only a finite number of measurements. Hence, to provide some perspective we also ran a simulation of a perfect device running this benchmark. The score of this ideal simulation indicates the best any real device can ever hope to get with these settings.

\paragraph{Quantum matrix inversion.}
In this benchmark, we calculate the total variation distance between the empirical probability distributions (i.e., histograms), and the ideal distributions that are proportional to the columns of the inverse matrix (with absolute value squared entries). The score of the benchmark is defined by these distances averaged over the different columns. For saving circuit depth we do not apply amplitude amplification, and therefore not all measurements give a ``valid'' outcome. In order to make the scoring fair, we take a weighted average over the total-variation distances weighed by the probabilities of obtaining ``valid'' outcomes. In order to weigh the results fairly, we consider the vector $v$ which contains the absolute value square of the matrix entries in $A^{-1}$, and the corresponding vector of (subnormalized) empirical distribution histograms $\tilde{v}$, and define the score as $\lVert v-\tilde{v} \rVert_1/(\lVert v \rVert_1 + \lVert \tilde{v} \rVert_1)$.

Every entry in the empirical probability distribution is determined by the fraction of measurement outcomes corresponding to that entry. Since subsequent measurements are independent, each of the entries itself is distributed according to a binomial distribution. The standard error follows from the properties of this distribution and standard error propagation. We remark that the entries within each column are not necessarily independent from one another, but the entries across different columns are.

We define the final overall score as the average of the scores of the first three instances (corresponding to different matrix dimensions) of this benchmark family.

\paragraph{Platonic Fractals.}
For $n$ measurements, the resulting circuit acts on $n+1$ qubits, contains between $1+4n$ and $1+8n$ gates, and has depth between $2+n$ and $2+5n$ (except for $n = 1$ where the minimal depth is $4$ rather than $3$).

For each point that we calculated from the measurement statistics, we calculate its $\ell^2$-distance to its reference point that can be calculated analytically (cf.\ \cref{eq:weak}). The score of this benchmark, now, is the average between all of these distances, and the standard error we estimate by taking the standard deviation of all the distances that we have obtained.

To all devices that allow for the execution of the first three stages of this benchmark, we associate the overall score, which we calculate by
\[
\mathrm{score} = \frac13(s_1+s_2+s_3) \pm \frac13\sqrt{\sigma_1^2 + \sigma_2^2 + \sigma_3^2}.
\]

\subsection{Score Error Margins}\label{sec:errors}
There are three distinct sources of variability in the benchmarks when run on real-world devices. The first one is measurement noise: for each of the benchmarks, multiple measurement outcomes are collated to obtain a counting statistic (i.e.\ a histogram) as an estimate of the real probability distribution representing the final state. The second source of errors are stochastic errors throughout the circuit: gates are not truly unitary but rather quantum channels, and depolarizing noise (as well as more complex noise sources) decohere the quantum state. The third set of errors stems from the variability of the device itself: external temperature fluctuations~\cite{kenneth2019ProbingContextDepErrors} and memory effects~\cite{maciejewski2021ModellAndMitigateReadoutNoiseQAOA} make gates ``drift'' away from their intended action, and, if the data is collected over a longer time span, then the device may be tuned up differently from time to time.
These errors all affect the various benchmark results, and hence the scores we assign to every test come with an uncertainty margin which is carefully chosen for each benchmark as we describe below.

The presented scores compare the obtained experimental results with the ideal statistics that we would get by repeating the experiment indefinitely on a perfect device. Ideally, we would like to repeat our experiments indefinitely to exactly determine the deviation compared to a perfect device, based on which we could compute the ``ideal score''. Since we can only afford a finite number of measurements, we can only give an estimate of this ``ideal score'', and even a perfect device would exhibit some statistical errors in our plots, cf. the simulated results. Often we chose a fine-tuned formula for the error margin so that it reflects the magnitude of this statistical error, i.e., the margins indicate how much of the estimated error score can be reasonably explained by statistical errors, i.e., ``measurement noise''. To see the usefulness of this approach, consider for example the Schrödinger's Microscope benchmark on a perfect (or simulated) device: if we use a few thousands of measurements, then the expected score becomes about $1\%$, while due to the high number of pixels displayed, the standard deviation of the numerical value of the score is minuscule. However, in this case our error margins display the ``expected deviation'' from the ideal score, and are in agreement with the error score estimates that we observe in simulated experiments. Meanwhile, in other cases when the collected results come in several well-defined batches, we present error margins that simply represent the standard deviation of the obtained numerical score values coming from the multiple batches, such as in the Bell-test and Line-Drawing benchmarks.
 
In the experimental data that we obtained we also see some fluctuations over time. The devices appear to have ``better'' and ``worse'' time-periods and the estimated scores might vary a lot in time. However, estimating this variance in time would require performing considerably more experiments, therefore our error margins only encompass statistical uncertainty and we leave the study of such temporal inhomogeneities to further works~\cite{dasgupta2020CharStabilityOfNISQDevs}.

\begin{figure}[p]
	\centering
	\vspace{-2em}
	\includegraphics[width=\textwidth]{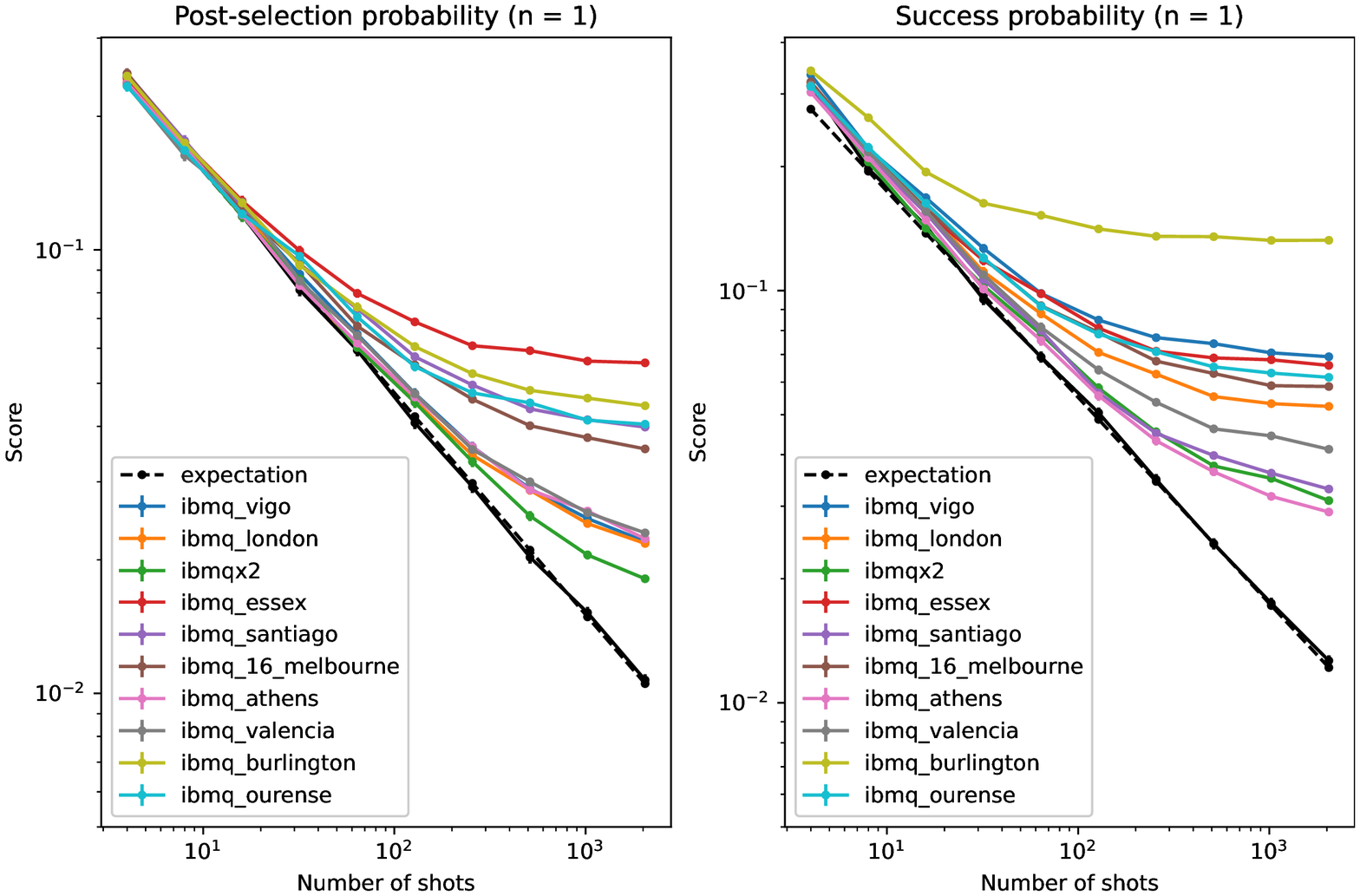}
	\includegraphics[width=\textwidth]{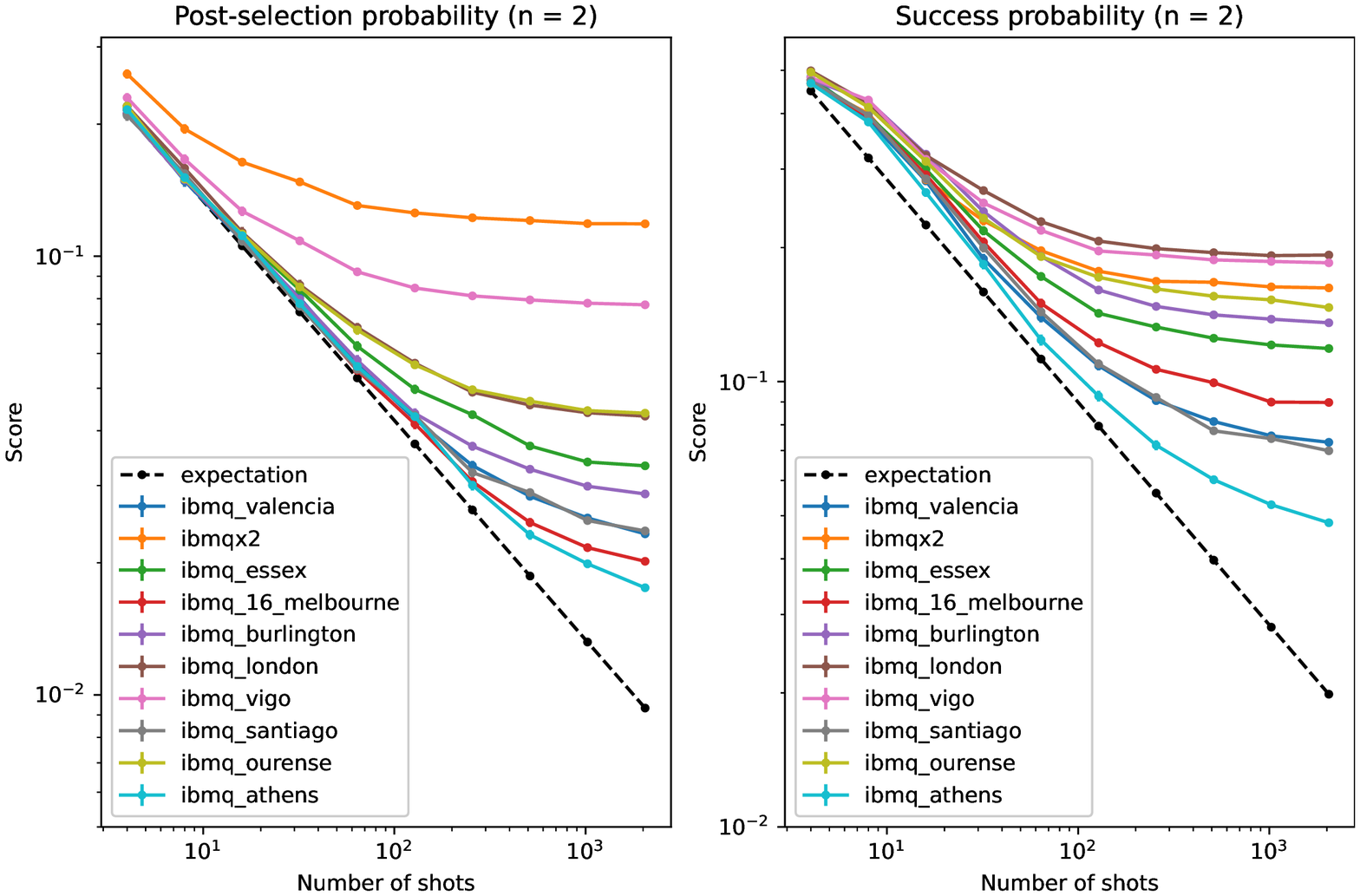}
	\caption{A plot of the achieved score versus the number of shots for every circuit in the Schr\"odinger-Microscope benchmark. It can be seen that for a small number of shots, the statistical noise dominates since the resulting score is close to the simulated score, but with an increasing number of shots, the device noise starts to dominate for all devices that we ran our benchmarks on.}
	\label{fig:extrapolation}
\end{figure}

\subsection{Disentangling Statistical and Device Noise}\label{sec:disentangling}
Disregarding fluctuations over time the two leftover sources for error within each benchmark are statistical and device noise.
The first arises from the fact that we are estimating the probability that a measurement outcome equals $1$ from finitely many samples. This noise is present in all runs, i.e., in the simulated runs as well as in the runs on the actual hardware.
The device noise is the error quantity we are actually interested in. Both contribute to the total noise in the experiment, which is what the score reflects.

In order to understand what part of the score is actually device noise, and what part is statistical noise, we note that the contribution of the statistical noise scales as $1/\sqrt{N}$, where $N$ is the number of shots, whereas the device noise is constant with respect to the number of runs that are performed.

We demonstrate this through the example of the Schr\"odinger-Microscope benchmark. We shuffle our $4096$ runs, and then divide them into several bins, first of size $2048$, then $1024$, then $512$, etc., with the smallest bin being of size $4$. With these bins, we calculate our score. We now expect that the score behaves as $n_s/\sqrt{N} + n_d$, where $n_s$ is the statistical noise and $n_d$ is the device noise. We plot this in \cref{fig:extrapolation}. The predicted behavior is indeed apparent, providing evidence for our hypothesis that the noise consists of these two major contributions.

\subsection{Device-Specific Results}
\label{subsec:results}
\newcommand{\qubit}{
circle[fill=white,radius=3pt]
}
\newcommand{\topo}[1]{\begin{tikzpicture}[
    scale=0.3
]
#1
\end{tikzpicture}}
\newcommand\topolinfive{
\raisebox{.4em}{\topo{\draw (0,0) \qubit -- (1,0) \qubit -- (2,0) \qubit -- (3,0) \qubit -- (4,0) \qubit;}}
}
\newcommand\topoTfive{
\topo{\draw (0,0) \qubit -- (1,0) \qubit -- (2,0) \qubit; \draw (1,0) -- (1,-1) \qubit -- (1,-2) \qubit;}
}
\newcommand\topostarfive{
\topo{\draw (1,1) \qubit -- (0,0) \qubit -- (1,0) \qubit -- (2,0) \qubit -- (1,-1) \qubit; \draw (1,1) -- (1,0) -- (1,-1);}
}
\newcommand\toposquarefive{\topo{
	\draw (0,0) \qubit -- (0,-1.4) \qubit -- (.7,-.7) \qubit -- (1.4,0) \qubit -- (1.4,-1.4) \qubit -- (.7,-.7) -- (0,0);
}}
\newcommand\toposidewaysH{\topo{
	\draw (0,0) \qubit -- (1,0) \qubit -- (2,0) \qubit;
	\draw (0,-2) \qubit -- (1,-2) \qubit -- (2,-2) \qubit;
	\draw (1,0) -- (1,-1) \qubit -- (1,-2);
}}
\newcommand\topomelbourne{\topo{
	\draw (0,0) \qubit -- (1,0) \qubit -- (2,0) \qubit -- (3,0) \qubit -- (4,0) \qubit -- (5,0) \qubit -- (6,0) \qubit -- (7,0) \qubit;
	\draw (0,-1) \qubit -- (1,-1) \qubit -- (2,-1) \qubit -- (3,-1) \qubit -- (4,-1) \qubit -- (5,-1) \qubit -- (6,-1) \qubit -- (7,-1) \qubit;
	\draw (0,0) -- (0,-1);
	\draw (1,0) -- (1,-1);
	\draw (2,0) -- (2,-1);
	\draw (3,0) -- (3,-1);
	\draw (4,0) -- (4,-1);
	\draw (5,0) -- (5,-1);
	\draw (6,0) -- (6,-1);
	\draw (7,0) -- (7,-1);
}}
\newcommand\aspenring{
	\draw (0,0) \qubit -- (0,1) \qubit -- (.7,1.7) \qubit -- (1.7,1.7) \qubit -- (2.4,1) \qubit -- (2.4,0) \qubit -- (1.7,-.7) \qubit -- (.7,-.7) \qubit -- (0,0);
}
\newcommand\topoaspenfour{\topo{
	\aspenring
	\draw (2.4,1) -- (3.4,1);
	\draw (2.4,0) -- (3.4,0);
	\begin{scope}[shift={(3.4,0)}]
		\aspenring
	\end{scope}
}}
\newcommand\topoaspenseven{\topo{
	\aspenring
	\draw (2.4,1) -- (3.4,1);
	\draw (2.4,0) -- (3.4,0);
	\begin{scope}[shift={(3.4,0)}]
		\aspenring
		\draw (2.4,1) -- (3.4,1);
		\draw (2.4,0) -- (3.4,0);
	\end{scope}
	\begin{scope}[shift={(6.8,0)}]
		\aspenring
		\draw (2.4,1) -- (3.4,1);
		\draw (2.4,0) -- (3.4,0);
	\end{scope}
	\begin{scope}[shift={(10.2,0)}]
		\aspenring
	\end{scope}
}}
\newcommand\topoionq{\topo{
	\foreach \x in {0,1,...,10} {
		\foreach \y in {0,1,...,\x} {
			\draw ({2*cos(\x/11*360)},{2*sin(\x/11*360)}) \qubit -- ({2*cos(\y/11*360)},{2*sin(\y/11*360)});
		}
	}
}}

\begin{table}[t!]
    \thisfloatpagestyle{empty}
	\centering
	\begin{tabular}{m{1.6cm} m{2cm} m{1.6cm} m{4cm} m{1cm} m{1.2cm}}
		\toprule
		Vendor & Name & \# qubits & Topology & QV & avg 2Q gate fid \\
		\midrule
		\multirow{20}{*}{IBM} & athens & $5$ & \topolinfive & 32 & 0.992(1) \\
		& belem & $5$ & \topoTfive & 16 & 0.99(0) \\
		& bogota & $5$ & \topolinfive & 32 & 0.98(7) \\
		& burlington & $5$ & \topoTfive & 8 & 0.9(5) \\
		& casablanca & $7$ & \toposidewaysH & 32 & 0.98(9) \\
		& essex & $5$ & \topoTfive & 8 & 0.99(0) \\
		& ibmqx2 & $5$ & \toposquarefive & 8 & 0.98(1) \\
		& lima & $5$ & \topoTfive & 8 & 0.98(8) \\
		& london & $5$ & \topoTfive & 16 & 0.98(7) \\
		& melbourne & $14$ / $15$* & \topomelbourne & 4/8$^\dagger$ & 0.9(7) \\
		& ourense & $5$ & \topoTfive & 8 & 0.99(2) \\
		& quito & $5$ & \topoTfive & 8 & 0.98(9) \\
		& rome & $5$ & \topolinfive & 32 & 0.98(8)  \\
		& santiago & $5$ & \topolinfive & 32 & 0.99(0) \\
		& valencia & $5$ & \topoTfive & 16 & 0.98(9) \\
		& vigo & $5$ & \topoTfive & 16  & 0.99(0) \\
		\midrule
		\multirow{7}{*}{Rigetti} & Aspen-4 & $13$* & \topoaspenfour & 8 & 0.976 \\
		& Aspen-7 & $28$* & \topoaspenseven & 8$^\ddagger$ & 0.757 \\
		& Aspen-8 & $31$* & \topoaspenseven & 8$^\ddagger$ & 0.933 \\
		& Aspen-9 & $31$* & \topoaspenseven & 8$^\ddagger$ & 0.871 \\
        \midrule
        Amazon & IonQ & $11$ & \topoionq & 32 & 0.970 \\
		\bottomrule
	\end{tabular}
	\caption{Benchmarked devices, their qubit counts, layouts, and quantum volume. The qubit counts marked with an asterisk do not match the number of qubits shown in the layout. In these cases, some of the physical qubits could not be accessed by thy user. Melbourne has been listed both with $14$ and $15$ accessible qubits.
	Quantum volume (QV) for melbourne ($\dagger$) is reported 8 on the IBM quantum experience, but has previously been measured to be 4. The Aspen QV values ($\ddagger$, apart from Aspen-4) are all estimated from their gate fidelities, as explained in \cref{sec:qv}.}
	\label{tbl:devices}
\end{table}

We have run our benchmarks on all the quantum devices listed in \cref{tbl:devices}. Below follow the resulting figures.

We display our findings using \textit{result cards}, one per device and benchmark. With all the scores that we list, we also supply a $1\sigma$-error margin (see \cref{sec:errors} for a discussion of how device and statistical noise are collated to this margin). A questionmark (?) denotes an unavailable score at the time of submission, due to a missing datapoint.

\clearpage
\storeareas\restoregeometry
\areaset{\textwidth+3cm}{\textheight+2.5cm}
\pagestyle{empty}

\DeclareDocumentCommand{\resultsCard}{m m m m}{%
\begin{tikzpicture}[
    baseline={([yshift={-\ht\strutbox}]current bounding box.north)}
]
    \draw[black!50,use as bounding box] (-.1,1.65) rectangle (4.4,-#4-.1);
    \draw[black!50,dashed] (0,1.55) rectangle (4.3,-#4);
    \node[anchor=north west] at (0, 1.5) {#1};
    \node[anchor=north west] at (0, .9) {\textbf{#2}};
    #3
\end{tikzpicture}
}

\clearpage
\subsubsection{Bell Test}
\DeclareDocumentCommand{\resultsCardImg}{m O{4cm}}{%
    \IfFileExists{#1}
    {\includegraphics[width=#2,draft=\hidepics]{#1}}
	{\parbox[c]{.2\textwidth}{\centering \vspace{1.3em} ~\\ $\;$ N/A}}
}
\DeclareDocumentCommand{\bellResultsCard}{m m m m}{%
\resultsCard{#1}{#2}{
    \node[anchor=north west] at (0, 0) {\resultsCardImg{experiments/bell/#2.pdf}};
    \node[anchor=north east,xshift=4mm,scale=1.3] at (4, 1.5)
    {$#3$};
    \node[anchor=north east,xshift=3mm,scale=.9,opacity=.7] at (4, .9)
    {#4};
}{4.3}
}

\noindent
\bellResultsCard{IBM}{athens}{1.38 {\color{gray}{\pm 0.03}}}{Oct 2020}
\bellResultsCard{IBM}{belem}{1.25 {\color{gray}{\pm 0.05}}}{Mar 2021}
\bellResultsCard{IBM}{bogota}{1.14 {\color{gray}{\pm 0.06}}}{Mar 2021}
\bellResultsCard{IBM}{burlington}{0.53 {\color{gray}{\pm 0.33}}}{Nov 2019}
\\[-5mm]

\noindent
\bellResultsCard{IBM}{casablanca}{1.21 {\color{gray}{\pm 0.09}}}{Mar 2021}
\bellResultsCard{IBM}{essex}{1.28 {\color{gray}{\pm 0.10}}}{Nov 2019}
\bellResultsCard{IBM}{ibmqx2}{1.33 {\color{gray}{\pm 0.10}}}{Nov 2019}
\bellResultsCard{IBM}{lima}{1.14 {\color{gray}{\pm 0.15}}}{Mar 2021}
\\[-5mm]

\noindent
\bellResultsCard{IBM}{london}{1.08 {\color{gray}{\pm 0.04}}}{Nov 2019}
\bellResultsCard{IBM}{melbourne}{0.85 {\color{gray}{\pm 0.29}}}{Sep 2020}
\bellResultsCard{IBM}{ourense}{1.21 {\color{gray}{\pm 0.22}}}{Nov 2019}
\bellResultsCard{IBM}{quito}{1.17 {\color{gray}{\pm 0.05}}}{Mar 2021}
\\[-5mm]

\noindent
\bellResultsCard{IBM}{rome}{1.08 {\color{gray}{\pm 0.13}}}{Mar 2021}
\bellResultsCard{IBM}{santiago}{1.36 {\color{gray}{\pm 0.05}}}{Sep 2020}
\bellResultsCard{IBM}{valencia}{1.26 {\color{gray}{\pm 0.04}}}{Sep 2020}
\bellResultsCard{IBM}{vigo}{1.33 {\color{gray}{\pm 0.06}}}{Nov 2019}
\\[-5mm]

\noindent
\bellResultsCard{Rigetti}{Aspen-4}{1.16 {\color{gray}{\pm 0.22}}}{Nov 2019}%
\bellResultsCard{Rigetti}{Aspen-8}{0.97 {\color{gray}{\pm 0.39}}}{Sep 2020}
\\

\vfill
\definecolor{gradient1}{rgb}{0.987622,0.64532,0.039886}
\definecolor{gradient2}{rgb}{0.729909,0.212759,0.333861}
\definecolor{gradient3}{rgb}{0.258234,0.03857,0.406485}
\begin{center}
\begin{tikzpicture}[x=1cm,y=1cm,scale=1.18]
    \draw[left color=black!25, right color=black!88,draw=none] (0, -0.25) rectangle (10, 0.25);
    \draw[left color=gradient1, right color=gradient3, middle color=gradient2, draw=none] (10, -0.25) rectangle (15, 0.25);
    \foreach \d/\s/\o in {
        athens/1.38/1,
        belem/1.25/-3,
        bogota \& lima/1.14/-3,
        burlington/0.53/2,
        casablanca \& ourense/1.21/0,
        essex/1.28/2,
        ibmqx2 \& vigo/1.33/0,
        london \& rome/1.08/0,
        melbourne/0.85/-2,
        quito/1.17/3,
        santiago/1.36/0,
        valencia/1.26/1,
        Aspen-4/1.16/-1,
        Aspen-8/0.97/0
    }{
        \draw[black] (10*\s,-0.25) -- (10*\s,0.25) -- (10*\s+.05,0.5) node[xshift=\o,rotate=75,anchor=west] {\d};
    }
    \foreach \x in {0.0, 0.5, 1.0, 1.5} {
        \draw (10*\x, -.4) -- (10*\x, -.6) node[anchor=north] {$\x$};
    }
    \draw[->] (9.2, -.5) -- (8.2, -.5) node[anchor=east] {classical};
    \draw[->] (10.8, -.5) -- (11.8, -.5) node[anchor=west] {quantum};
\end{tikzpicture}
\end{center}

\clearpage
\subsubsection{Schr\"odinger's Microscope}
\DeclareDocumentCommand{\SMSmallResultsCard}{m m m m m m m m}{%
	\resultsCard{#1}{#2}{
		\node[anchor=north west] at (0, .5) {\resultsCardImg{experiments/SM/#2.1.pdf}};
		\node at (1.1,-1.75) {\scriptsize $#3$};
		\node at (3.2,-1.75) {\scriptsize $#4$};
		\node[anchor=north west] at (0, -2) {\resultsCardImg{experiments/SM/#2.2.pdf}};
		\node at (1.1,-4.25) {\scriptsize $#5$};
		\node at (3.2,-4.25) {\scriptsize $#6$};
		\node[anchor=north east,xshift=3mm] at (4, 1.5)
		{$#8$};
		\node[anchor=north east,xshift=2mm,scale=.8,opacity=.7] at (4, .9)
		{#7};
	}{4.5}
}

\newcommand\SMLargeResultsCard[2]{%
	\def\vendor{#1}%
	\def\device{#2}%
	\SMLargeResultsCardTmp
}

\DeclareDocumentCommand{\SMLargeResultsCardTmp}{m m m m m m m m}{%
    \resultsCard{\vendor}{\device}{
		\node[anchor=north west] at (0, .5) {\resultsCardImg{experiments/SM/\device.1.pdf}};
		\node at (1.1,-1.75) {\scriptsize $#1$};
		\node at (3.2,-1.75) {\scriptsize $#2$};
		\node[anchor=north west] at (0, -2) {\resultsCardImg{experiments/SM/\device.2.pdf}};
		\node at (1.1,-4.25) {\scriptsize $#3$};
		\node at (3.2,-4.25) {\scriptsize $#4$};
		\node[anchor=north west] at (0, -4.5) {\resultsCardImg{experiments/SM/\device.3.pdf}};
		\node at (1.1,-6.75) {\scriptsize $#5$};
		\node at (3.2,-6.75) {\scriptsize $#6$};
		\node[anchor=north east,xshift=3mm] at (4, 1.5)
		{$#8$};
		\node[anchor=north east,xshift=2mm,scale=.8,opacity=.7] at (4, .9)
		{#7};
	}{7}
}
\enlargethispage{3cm}

\noindent
\resizebox{.25\textwidth}{!}{\SMLargeResultsCard{-}{simulation}{0.0073 {\color{gray}{\pm 0.0002}}}{0.0082 {\color{gray}{\pm 0.0003}}}{0.0057 {\color{gray}{\pm 0.0002}}}{0.0145 {\color{gray}{\pm 0.0005}}}{0.0054 {\color{gray}{\pm 0.0002}}}{0.0272 {\color{gray}{\pm 0.0011}}}{Nov 2020}{0.0089 {\color{gray}{\pm 0.0002}}}}
\resizebox{.25\textwidth}{!}{\SMLargeResultsCard{IBM}{melbourne}{0.0349 {\color{gray}{\pm 0.0002}}}{0.0577 {\color{gray}{\pm 0.0003}}}{0.0494 {\color{gray}{\pm 0.0002}}}{0.1550 {\color{gray}{\pm 0.0005}}}{0.0536 {\color{gray}{\pm 0.0002}}}{0.2630 {\color{gray}{\pm 0.0012}}}{Nov 2019}{0.0742 {\color{gray}{\pm 0.0002}}}}
\resizebox{.25\textwidth}{!}{\SMLargeResultsCard{Rigetti}{Aspen-4}{0.0276 {\color{gray}{\pm 0.0002}}}{0.0584 {\color{gray}{\pm 0.0003}}}{?}{?}{0.1139 {\color{gray}{\pm 0.0003}}}{0.3042 {\color{gray}{\pm 0.0019}}}{Oct 2019}{?}}
\resizebox{.25\textwidth}{!}{\SMLargeResultsCard{Rigetti}{Aspen-7}{0.0622 {\color{gray}{\pm 0.0002}}}{0.1451 {\color{gray}{\pm 0.0003}}}{?}{?}{0.0971 {\color{gray}{\pm 0.0003}}}{0.2900 {\color{gray}{\pm 0.0017}}}{Nov 2019}{?}}
\\[5mm]

\noindent
\resizebox{.25\textwidth}{!}{\SMSmallResultsCard{IBM}{athens}{0.0214 {\color{gray}{\pm 0.0002}}}{0.0276 {\color{gray}{\pm 0.0003}}}{0.0094 {\color{gray}{\pm 0.0002}}}{0.0369 {\color{gray}{\pm 0.0005}}}{Nov 2020}{0.0238 {\color{gray}{\pm 0.0001}}}}
\resizebox{.25\textwidth}{!}{\SMSmallResultsCard{IBM}{belem}{0.0279 {\color{gray}{\pm 0.0002}}}{0.1236 {\color{gray}{\pm 0.0003}}}{0.0306 {\color{gray}{\pm 0.0002}}}{0.1082 {\color{gray}{\pm 0.0005}}}{Mar 2021}{0.0726 {\color{gray}{\pm 0.0002}}}}
\resizebox{.25\textwidth}{!}{\SMSmallResultsCard{IBM}{bogota}{0.0244 {\color{gray}{\pm 0.0002}}}{0.0411 {\color{gray}{\pm 0.0003}}}{0.0140 {\color{gray}{\pm 0.0002}}}{0.0665 {\color{gray}{\pm 0.0005}}}{Mar 2021}{0.0365 {\color{gray}{\pm 0.0003}}}}
\resizebox{.25\textwidth}{!}{\SMSmallResultsCard{IBM}{burlington}{0.0440 {\color{gray}{\pm 0.0002}}}{0.1319 {\color{gray}{\pm 0.0003}}}{0.0242 {\color{gray}{\pm 0.0002}}}{0.1350 {\color{gray}{\pm 0.0005}}}{Nov 2019}{0.0838 {\color{gray}{\pm 0.0002}}}}
\\[5mm]

\noindent
\resizebox{.25\textwidth}{!}{\SMSmallResultsCard{IBM}{casablanca}{0.0241 {\color{gray}{\pm 0.0002}}}{0.0328 {\color{gray}{\pm 0.0003}}}{0.0126 {\color{gray}{\pm 0.0002}}}{0.0576 {\color{gray}{\pm 0005}}}{Mar 2021}{0.0318 {\color{gray}{\pm 0.0003}}}}
\resizebox{.25\textwidth}{!}{\SMSmallResultsCard{IBM}{essex}{0.0551 {\color{gray}{\pm 0.0002}}}{0.0653 {\color{gray}{\pm 0.0003}}}{0.0290 {\color{gray}{\pm 0.0002}}}{0.1183 {\color{gray}{\pm 0.0005}}}{Nov 2019}{0.0669 {\color{gray}{\pm 0.0002}}}}
\resizebox{.25\textwidth}{!}{\SMSmallResultsCard{IBM}{ibmqx2}{0.0164 {\color{gray}{\pm 0.0002}}}{0.0304 {\color{gray}{\pm 0.0003}}}{0.0181 {\color{gray}{\pm 0.0002}}}{0.0638 {\color{gray}{\pm 0.0005}}}{Nov 2019}{0.0322 {\color{gray}{\pm 0.0002}}}}
\resizebox{.25\textwidth}{!}{\SMSmallResultsCard{IBM}{lima}{0.0176 {\color{gray}{\pm 0.0002}}}{0.0258 {\color{gray}{\pm 0.0003}}}{0.0193 {\color{gray}{\pm 0.0002}}}{0.0644 {\color{gray}{\pm 0.0005}}}{Feb 2021}{0.0318 {\color{gray}{\pm 0.0002}}}}
\\[5mm]

\noindent
\resizebox{.25\textwidth}{!}{\SMSmallResultsCard{IBM}{london}{0.0206 {\color{gray}{\pm 0.0002}}}{0.0513 {\color{gray}{\pm 0.0003}}}{0.0481 {\color{gray}{\pm 0.0002}}}{0.1908 {\color{gray}{\pm 0.0058}}}{Oct 2019}{0.0777 {\color{gray}{\pm 0.0014}}}}
\resizebox{.25\textwidth}{!}{\SMSmallResultsCard{IBM}{ourense}{0.0397 {\color{gray}{\pm 0.0002}}}{0.0611 {\color{gray}{\pm 0.0003}}}{0.0320 {\color{gray}{\pm 0.0002}}}{0.1481 {\color{gray}{\pm 0.0005}}}{Nov 2019}{0.0702 {\color{gray}{\pm 0.0002}}}}
\resizebox{.25\textwidth}{!}{\SMSmallResultsCard{IBM}{quito}{0.0378 {\color{gray}{\pm 0.0002}}}{0.0609 {\color{gray}{\pm 0.0058}}}{0.0168 {\color{gray}{\pm 0.0002}}}{0.0638 {\color{gray}{\pm 0.0005}}}{Feb 2021}{0.0448 {\color{gray}{\pm 0.0014}}}}
\resizebox{.25\textwidth}{!}{\SMSmallResultsCard{IBM}{rome}{0.0299 {\color{gray}{\pm 0.0002}}}{0.0675 {\color{gray}{\pm 0.0005}}}{0.0191 {\color{gray}{\pm 0.0002}}}{0.0352 {\color{gray}{\pm 0.0003}}}{Mar 2021}{0.0379 {\color{gray}{\pm 0.0003}}}}
\\[5mm]

\noindent
\resizebox{.25\textwidth}{!}{\SMSmallResultsCard{IBM}{santiago}{0.0395 {\color{gray}{\pm 0.0002}}}{0.0318 {\color{gray}{\pm 0.0003}}}{0.0201 {\color{gray}{\pm 0.0001}}}{0.0670 {\color{gray}{\pm 0.0004}}}{Oct 2020}{0.0494 {\color{gray}{\pm 0.0002}}}}
\resizebox{.25\textwidth}{!}{\SMSmallResultsCard{IBM}{valencia}{0.0221 {\color{gray}{\pm 0.0002}}}{0.0407 {\color{gray}{\pm 0.0003}}}{0.0183 {\color{gray}{\pm 0.0002}}}{0.0561 {\color{gray}{\pm 0.0004}}}{Sep 2020}{0.0343 {\color{gray}{\pm 0.0001}}}}
\resizebox{.25\textwidth}{!}{\SMSmallResultsCard{IBM}{vigo}{0.0208 {\color{gray}{\pm 0.0002}}}{0.0689 {\color{gray}{\pm 0.0003}}}{0.0239 {\color{gray}{\pm 0.0002}}}{0.1128 {\color{gray}{\pm 0.0005}}}{Oct 2019}{0.0566 {\color{gray}{\pm 0.0002}}}}
\resizebox{.25\textwidth}{!}{\SMSmallResultsCard{Rigetti}{Aspen-8}{0.1160 {\color{gray}{\pm 0.0002}}}{0.0809 {\color{gray}{\pm 0.0003}}}{0.0558 {\color{gray}{\pm 0.0002}}}{0.1809 {\color{gray}{\pm 0.0006}}}{Oct 2020}{0.1084 {\color{gray}{\pm 0.0002}}}}

\clearpage
\subsubsection{Mandelbrot}
\DeclareDocumentCommand{\MandelbrotSuperSmallResultsCard}{m m m m m m}{%
	\resultsCard{#1}{#2}{
		\node[anchor=north west] at (0, .5) {\resultsCardImg{experiments/Mandelbrot/#2.1.pdf}};
		\node at (1.1,-1.75) {\scriptsize $#3$};
		\node at (3.2,-1.75) {\scriptsize $#4$};
		\node[anchor=north east,xshift=3mm] at (4, 1.5)
		{$#6$};
		\node[anchor=north east,xshift=2mm,scale=.8,opacity=.7] at (4, .9)
		{#5};
	}{2}
}

\DeclareDocumentCommand{\MandelbrotSmallResultsCard}{m m m m m m m m}{%
	\resultsCard{#1}{#2}{
		\node[anchor=north west] at (0, .5) {\resultsCardImg{experiments/Mandelbrot/#2.1.pdf}};
		\node at (1.1,-1.75) {\scriptsize $#3$};
		\node at (3.2,-1.75) {\scriptsize $#4$};
		\node[anchor=north west] at (0, -2) {\resultsCardImg{experiments/Mandelbrot/#2.2.pdf}};
		\node at (1.1,-4.25) {\scriptsize $#5$};
		\node at (3.2,-4.25) {\scriptsize $#6$};
		\node[anchor=north east,xshift=3mm] at (4, 1.5)
		{$#8$};
		\node[anchor=north east,xshift=2mm,scale=.8,opacity=.7] at (4, .9)
		{#7};
	}{4.5}
}

\newcommand\MandelbrotLargeResultsCard[2]{%
	\def\vendor{#1}%
	\def\device{#2}%
	\MandelbrotLargeResultsCardTmp
}

\DeclareDocumentCommand{\MandelbrotLargeResultsCard}{m m m m m m m m}{%
	\resultsCard{\vendor}{\device}{
		\node[anchor=north west] at (0, .5) {\resultsCardImg{experiments/Mandelbrot/\device.1.pdf}};
		\node at (1.1,-1.75) {\scriptsize $#1$};
		\node at (3.2,-1.75) {\scriptsize $#2$};
		\node[anchor=north west] at (0, -2) {\resultsCardImg{experiments/Mandelbrot/\device.2.pdf}};
		\node at (1.1,-4.25) {\scriptsize $#3$};
		\node at (3.2,-4.25) {\scriptsize $#4$};
		\node[anchor=north west] at (0, -4.5) {\resultsCardImg{experiments/Mandelbrot/\device.3.pdf}};
		\node at (1.1,-6.75) {\scriptsize $#5$};
		\node at (3.2,-6.75) {\scriptsize $#6$};
		\node[anchor=north east,xshift=3mm] at (4, 1.5)
		{$#8$};
		\node[anchor=north east,xshift=2mm,scale=.8,opacity=.7] at (4, .9)
		{#7};
	}{7}
}

\enlargethispage{3cm}

\noindent
\MandelbrotSmallResultsCard{-}{simulation}{0.0060 {\color{gray}{\pm 0.0002}}}{0.0075 {\color{gray}{\pm 0.0002}}}{0.0071 {\color{gray}{\pm 0.0002}}}{0.0189 {\color{gray}{\pm 0.0010}}}{Nov 2020}{0.0099 {\color{gray}{\pm 0.0003}}}
\MandelbrotSmallResultsCard{IBM}{athens}{0.0240 {\color{gray}{\pm 0.0002}}}{0.0244 {\color{gray}{\pm 0.0002}}}{0.0599 {\color{gray}{\pm 0.0002}}}{0.1659 {\color{gray}{\pm 0.0009}}}{Oct 2020}{0.0685 {\color{gray}{\pm 0.0002}}}
\MandelbrotSmallResultsCard{IBM}{belem}{0.0782 {\color{gray}{\pm 0.0002}}}{0.1006 {\color{gray}{\pm 0.0003}}}{0.0931 {\color{gray}{\pm 0.0002}}}{0.2627 {\color{gray}{\pm 0.0006}}}{Mar 2021}{0.1337 {\color{gray}{\pm 0.0002}}}
\MandelbrotSmallResultsCard{IBM}{bogota}{0.0625{\color{gray}{\pm 0.0002}}}{0.0349 {\color{gray}{\pm 0.0003}}}{0.0789 {\color{gray}{\pm 0.0002}}}{0.2170 {\color{gray}{\pm 0.0007}}}{Mar 2021}{0.0983 {\color{gray}{\pm 0.0002}}}
\\[1mm]

\noindent
\MandelbrotSmallResultsCard{IBM}{burlington}{0.0798 {\color{gray}{\pm 0.0002}}}{0.0946 {\color{gray}{\pm 0.0003}}}{0.1126 {\color{gray}{\pm 0.0002}}}{0.3276 {\color{gray}{\pm 0.0006}}}{Jan 2020}{0.1537 {\color{gray}{\pm 0.0002}}}
\MandelbrotSmallResultsCard{IBM}{casablanca}{0.0733 {\color{gray}{\pm 0.0002}}}{0.0497 {\color{gray}{\pm 0.0003}}}{0.1042 {\color{gray}{\pm 0.0002}}}{0.3756 {\color{gray}{\pm 0.0005}}}{Apr 2021}{0.1507 {\color{gray}{\pm 0.0002}}}
\MandelbrotSmallResultsCard{IBM}{essex}{0.0426 {\color{gray}{\pm 0.0002}}}{0.0405 {\color{gray}{\pm 0.0002}}}{0.0897 {\color{gray}{\pm 0.0002}}}{0.2425 {\color{gray}{\pm 0.0006}}}{Jul 2020}{0.1038 {\color{gray}{\pm 0.0002}}}
\MandelbrotSmallResultsCard{IBM}{ibmqx2}{0.0338 {\color{gray}{\pm 0.0002}}}{0.0640 {\color{gray}{\pm 0.0003}}}{0.0830 {\color{gray}{\pm 0.0002}}}{0.2365 {\color{gray}{\pm 0.0007}}}{Jul 2020}{0.1043 {\color{gray}{\pm 0.0002}}}
\\[1mm]

\noindent
\MandelbrotSmallResultsCard{IBM}{lima}{0.0163 {\color{gray}{\pm 0.0002}}}{0.0328 {\color{gray}{\pm 0.0002}}}{0.0914 {\color{gray}{\pm 0.0002}}}{0.2216 {\color{gray}{\pm 0.0006}}}{Feb 2021}{0.0905 {\color{gray}{\pm 0.0002}}}
\MandelbrotSmallResultsCard{IBM}{london}{0.0169 {\color{gray}{\pm 0.0002}}}{0.0362 {\color{gray}{\pm 0.0002}}}{0.0796 {\color{gray}{\pm 0.0002}}}{0.2547 {\color{gray}{\pm 0.0007}}}{Jul 2020}{0.0969 {\color{gray}{\pm 0.0002}}}
\MandelbrotSmallResultsCard{IBM}{melbourne}{0.0404 {\color{gray}{\pm 0.0002}}}{0.0384 {\color{gray}{\pm 0.0002}}}{0.1160 {\color{gray}{\pm 0.0002}}}{0.2847 {\color{gray}{\pm 0.0006}}}{Jul 2020}{0.1199 {\color{gray}{\pm 0.0002}}}
\MandelbrotSmallResultsCard{IBM}{ourense}{0.0487 {\color{gray}{\pm 0.0002}}}{0.0533 {\color{gray}{\pm 0.0003}}}{0.0807 {\color{gray}{\pm 0.0002}}}{0.2902 {\color{gray}{\pm 0.0007}}}{Jul 2020}{0.1182 {\color{gray}{\pm 0.0002}}}
\\[1mm]

\noindent
\MandelbrotSmallResultsCard{IBM}{quito}{0.0393 {\color{gray}{\pm 0.0002}}}{0.0454 {\color{gray}{\pm 0.0003}}}{0.0955 {\color{gray}{\pm 0.0002}}}{0.2380 {\color{gray}{\pm 0.0006}}}{Feb 2021}{0.1046 {\color{gray}{\pm 0.0002}}}
\MandelbrotSmallResultsCard{IBM}{rome}{0.0206 {\color{gray}{\pm 0.0002}}}{0.0380 {\color{gray}{\pm 0.0002}}}{0.0915 {\color{gray}{\pm 0.0002}}}{0.2261 {\color{gray}{\pm 0.0007}}}{Mar 2021}{0.0941 {\color{gray}{\pm 0.0002}}}
\MandelbrotSmallResultsCard{IBM}{santiago}{0.0336 {\color{gray}{\pm 0.0002}}}{0.0179 {\color{gray}{\pm 0.0002}}}{0.0749 {\color{gray}{\pm 0.0002}}}{0.3249 {\color{gray}{\pm 0.0007}}}{Sep 2020}{0.1128 {\color{gray}{\pm 0.0002}}}
\MandelbrotSmallResultsCard{IBM}{valencia}{0.0224 {\color{gray}{\pm 0.0002}}}{0.0304 {\color{gray}{\pm 0.0002}}}{0.0886 {\color{gray}{\pm 0.0002}}}{0.2020 {\color{gray}{\pm 0.0006}}}{Sep 2020}{0.0859 {\color{gray}{\pm 0.0002}}}
\\[1mm]

\noindent
\MandelbrotSmallResultsCard{IBM}{vigo}{0.0327 {\color{gray}{\pm 0.0002}}}{0.0395 {\color{gray}{\pm 0.0002}}}{0.0695 {\color{gray}{\pm 0.0002}}}{0.1882 {\color{gray}{\pm 0.0007}}}{Jul 2020}{0.0825 {\color{gray}{\pm 0.0002}}}%
\MandelbrotSuperSmallResultsCard{Rigetti}{Aspen-7}{0.1954 {\color{gray}{\pm 0.0002}}}{0.1370 {\color{gray}{\pm 0.0003}}}{Nov 2019}{?}%
\MandelbrotSuperSmallResultsCard{Rigetti}{Aspen-8}{0.4232 {\color{gray}{\pm 0.0002}}}{0.3016 {\color{gray}{\pm 0.0004}}}{Jul 2020}{?}
\hfill~

\clearpage
\subsubsection{Line Drawing}
\DeclareDocumentCommand{\FirstArg}{m m}{#1}
\DeclareDocumentCommand{\SecondArg}{m m}{#2}
\DeclareDocumentCommand{\ScoreTwoLines}{ O{anchor=north east,align=right} m >{\SplitArgument{1}{+}}m O{1.3} O{0.9} O{-18} O{-1}}{%
	\node[#1,inner sep=0,outer sep=0,scale=#4] at (#2) { $\FirstArg #3 $ };
	\node[#1,inner sep=0,outer sep=0,yshift=#6,xshift=#7,scale=#5,opacity=.7] at (#2) { $\pm\SecondArg #3 $ };
}
\DeclareDocumentCommand{\ScoreTwoLinesSmall}{ O{anchor=north west,align=left} m m }{%
	\ScoreTwoLines[#1]{#2}{#3}[1.][.7][-12]
}
\DeclareDocumentCommand{\lineResultsCard}{
	m m
	m m
	o o o
}{%
	\resultsCard{#1}{#2}{
		\node[anchor=north west] at (0, .5) {\resultsCardImg{experiments/line/#2.4.pdf}};
		\IfValueTF{#5}{\ScoreTwoLinesSmall{0.1, .1}{#5}}{};
		\node[anchor=north west] at (0, -3) {\resultsCardImg{experiments/line/#2.8.pdf}};
		\IfValueTF{#6}{\ScoreTwoLinesSmall{0.1, -3.4}{#6}}{};
		\node[anchor=north west] at (0, -6.5) {\resultsCardImg{experiments/line/#2.16.pdf}};
		\IfValueTF{#7}{\ScoreTwoLinesSmall{0.1, -6.9}{#7}}{};
		\ScoreTwoLines{4.3, 1.4}{#3}
		\node[anchor=south west,align=left,opacity=.7] at (0, -11) {#4};
	}{11}
}

\DeclareDocumentCommand{\lineSmallResultsCard}{
	m m
	m m
	o o
}{%
	\resultsCard{#1}{#2}{
		\node[anchor=north west] at (0, .5) {\resultsCardImg{experiments/line/#2.4.pdf}};
		\IfValueTF{#5}{\ScoreTwoLinesSmall{0.1, .1}{#5}}{};
		\node[anchor=north west] at (0, -3) {\resultsCardImg{experiments/line/#2.8.pdf}};
		\IfValueTF{#6}{\ScoreTwoLinesSmall{0.1, -3.4}{#6}}{};
		\ScoreTwoLines{4.3, 1.4}{#3}
		\node[anchor=south west,align=left,opacity=.7] at (0, -7.5) {#4};
	}{7.5}
}

\noindent
\lineResultsCard{-}{simulation}{0.01+0.00}{Nov 2020}[0.01+0.00][0.01+0.00][0.01+0.00]
\lineResultsCard{IBM}{athens}{0.22+0.01}{Oct 2020}[0.05+0.01][0.20+0.02][0.41+0.03]
\lineResultsCard{IBM}{belem}{0.45+0.03}{Mar 2021}[0.15+0.07][0.38+0.04][0.81+0.03]
\lineResultsCard{IBM}{bogota}{0.40+0.02}{Mar 2021}[0.15+0.02][0.36+0.04][0.69+0.05]
\\[-5mm]

\enlargethispage{5cm}
\noindent
\lineResultsCard{IBM}{burlington}{0.47+0.03}{Dec 2019}[0.12+0.02][0.49+0.04][0.81+0.06]
\lineResultsCard{IBM}{casablanca}{0.50+0.06}{Mar 2021}[0.21+0.11][0.40+0.17][0.64+0.07]
\lineResultsCard{IBM}{essex}{0.40+0.02}{Dec 2019}[0.13+0.01][0.42+0.04][0.67+0.03]
\lineResultsCard{IBM}{ibmqx2}{0.49+0.02}{Jan 2020}[0.17+0.04][0.49+0.01][0.82+0.04]
\\[-5mm]

\noindent
\lineResultsCard{IBM}{lima}{0.37+0.03}{Mar 2021}[0.17+0.06][0.25+0.04][0.70+0.03]
\lineResultsCard{IBM}{london}{0.40+0.03}{Jul 2020}[0.12+0.01][0.36+0.09][0.73+0.04]
\lineResultsCard{IBM}{melbourne}{0.30+0.01}{Dec 2019}[0.10+0.01][0.24+0.03][0.59+0.02]
\lineResultsCard{IBM}{ourense}{0.25+0.02}{Jan 2020}[0.09+0.01][0.23+0.05][0.42+0.03]
\\[-5mm]

\enlargethispage{5cm}
\noindent
\lineResultsCard{IBM}{quito}{0.43+0.04}{Mar 2021}[0.22+0.09][0.39+0.06][0.69+0.06]
\lineResultsCard{IBM}{rome}{0.43+0.05}{Mar 2021}[0.13+0.02][0.40+0.13][0.76+0.04]
\lineResultsCard{IBM}{santiago}{0.28+0.06}{Sep 2020}[0.09+0.02][0.21+0.02][0.54+0.19]
\lineResultsCard{IBM \href{https://www.youtube.com/watch?v=qbGPXLQZ52g}{\ExternalLink}}{valencia}{0.32+0.03}{Sep 2020}[0.07+0.01][0.30+0.04][0.59+0.06]
\\[-5mm]

\noindent
\lineResultsCard{IBM}{vigo}{0.27+0.01}{Dec 2019}[0.13+0.01][0.25+0.02][0.45+0.02]%
\lineSmallResultsCard{Rigetti}{Aspen-8}{?+}{Jul 2020}[0.55+0.13][0.90+0.08]

\clearpage
\subsubsection{HHL}

\newcommand{\calcSc}[1]{\fpeval{round((#1)/3,2)}}
\newcommand{\calcUc}[1]{\fpeval{round(sqrt(#1)/3,2)}}
\DeclareDocumentCommand{\ScoreTwoLinesSmallHHL}{ O{anchor=north west,align=left} m m }{%
	\ScoreTwoLines[#1]{#2}{#3}[1.][.7][-2.4][25]
}

\DeclareDocumentCommand{\HHLSmallResultsCard}{m m >{\SplitArgument{1}{+}}m m >{\SplitArgument{1}{+}}O{N/A+N/A} >{\SplitArgument{1}{+}}O{N/A+N/A} >{\SplitArgument{1}{+}}O{N/A+N/A}}{%
	\resultsCard{#1}{#2}{
		\node[anchor=north east,xshift=3mm] at (4, 1.5)
		{{\FirstArg #3} {\scriptsize\color{gray}{$\pm {\SecondArg #3}$}}};			
		\node[anchor=north east,xshift=2mm,scale=.8,opacity=.7] at (4, .8)
		{#4};			
		\begin{scope}[shift={(0,0)}]
			\node[anchor=north west] at (.35, .5)				
			{\resultsCardImg{experiments/HHL/#2.2qubit-1ancilla-CZ.pdf}[33mm]};
			\node at (2.15,-3.1) {{\FirstArg #5} {\scriptsize\color{gray}{$\pm {\SecondArg #5}$}}};
		\end{scope}	
		\begin{scope}[shift={(0,-3.85)}]
			\node[anchor=north west] at (.35, .5)				
			{\resultsCardImg{experiments/HHL/#2.3qubit-1ancilla-CZ.pdf}[33mm]};
			\node at (2.15,-3.1) {{\FirstArg #6} {\scriptsize\color{gray}{$\pm {\SecondArg #6}$}}};
		\end{scope}
		\begin{scope}[shift={(0,-7.7)}]
			\node[anchor=north west] at (.35, .5)				
			{\resultsCardImg{experiments/HHL/#2.4qubit-1ancilla-CZ.pdf}[33mm]};
			\node at (2.15,-3.1) {{\FirstArg #7} {\scriptsize\color{gray}{$\pm {\SecondArg #7}$}}};
		\end{scope}
	}{11}
}

\enlargethispage{5cm}

\noindent
\HHLSmallResultsCard{-}{simulation}{\calcSc{0.005+0.005+0.017}+\calcUc{0.007^2+0.008^2+0.022^2}}{Feb 2021}[0.005+0.007][0.005+0.008][0.017+0.022]
\HHLSmallResultsCard{IonQ/Amazon}{IonQ}{\calcSc{0.070+0.149+0.126}+\calcUc{0.008^2+0.009^2+0.022^2}}{Feb 2021}[0.070+0.008][0.149+0.009][0.126+0.022]
\HHLSmallResultsCard{IBM}{athens}{\calcSc{0.125+0.242+0.203}+\calcUc{0.008^2+0.010^2+0.022^2}}{Nov 2020}[0.125+0.008][0.242+0.010][0.203+0.022]
\HHLSmallResultsCard{IBM}{belem}{\calcSc{0.260+0.347+0.343}+\calcUc{0.009^2+0.010^2+0.023^2}}{Mar 2021}[0.260+0.009][0.347+0.010][0.343+0.023]
\\[-5mm]

\noindent
\HHLSmallResultsCard{IBM}{bogota}{\calcSc{0.149+0.688+0.309}+\calcUc{0.009^2+0.010^2+0.023^2}}{Mar 2021}[0.149+0.009][0.688+0.010][0.309+0.023]
\HHLSmallResultsCard{IBM}{casablanca}{\calcSc{0.110+0.178+0.167}+\calcUc{0.008^2+0.010^2+0.022^2}}{Mar 2021}[0.110+0.008][0.178+0.010][0.167+0.022]
\HHLSmallResultsCard{IBM}{ibmqx2}{\calcSc{0.537+0.520+0.437}+\calcUc{0.009^2+0.011^2+0.022^2}}{Nov 2020}[0.537+0.009][0.520+0.011][0.437+0.022]
\HHLSmallResultsCard{IBM}{lima}{\calcSc{0.131+0.344+0.332}+\calcUc{0.009^2+0.011^2+0.023^2}}{Apr 2021}[0.131+0.009][0.344+0.011][0.332+0.023]
\\[-5mm]

\clearpage
\enlargethispage{5cm}

\noindent
\HHLSmallResultsCard{IBM}{melbourne}{\calcSc{0.197+0.565+0.341}+\calcUc{0.008^2+0.010^2+0.023^2}}{Nov 2020}[0.197+0.008][0.565+0.010][0.341+0.023] \HHLSmallResultsCard{IBM}{ourense}{\calcSc{0.146+0.288+0.307}+\calcUc{0.008^2+0.010^2+0.023^2}}{Nov 2020}[0.146+0.008][0.288+0.010][0.307+0.023]
\HHLSmallResultsCard{IBM}{quito}{\calcSc{0.308+0.197+0.356}+\calcUc{0.009^2+0.010^2+0.022^2}}{Apr 2021}[0.308+0.009][0.197+0.010][0.356+0.022]
\HHLSmallResultsCard{IBM}{rome}{\calcSc{0.201+0.482+0.257}+\calcUc{0.009^2+0.011^2+0.023^2}}{Mar 2021}[0.201+0.009][0.482+0.011][0.257+0.023]
\\[-5mm]

\noindent
\HHLSmallResultsCard{IBM}{santiago}{\calcSc{0.132+0.474+0.287}+\calcUc{0.009^2+0.010^2+0.023^2}}{Nov 2020}[0.132+0.009][0.474+0.010][0.287+0.023]
\HHLSmallResultsCard{IBM}{valencia}{\calcSc{0.272+0.139+0.272}+\calcUc{0.009^2+0.009^2+0.023^2}}{Nov 2020}[0.272+0.009][0.139+0.009][0.272+0.023]
\HHLSmallResultsCard{IBM}{vigo}{\calcSc{0.176+0.245+0.331}+\calcUc{0.009^2+0.010^2+0.023^2}}{Nov 2020}[0.176+0.009][0.245+0.010][0.331+0.023]
\HHLSmallResultsCard{Rigetti}{Aspen-8}{\calcSc{0.402+0.631+0.314}+\calcUc{0.009^2+0.009^2+0.023^2}}{Nov 2020}[0.402+0.009][0.631+0.009][0.314+0.023]
\\[-5mm]

\clearpage
\enlargethispage{5cm}

\noindent
\HHLSmallResultsCard{Rigetti/Amazon}{Aspen-9}{\calcSc{0.728+0.500+0.348}+\calcSc{0.008+0.010+0.023}}{Feb 2021}[0.728+0.008][0.500+0.010][0.348+0.023]\\[-5mm]
\clearpage

\DeclareDocumentCommand{\HHLSmallResultsCardNext}{m m >{\SplitArgument{1}{+}}m m >{\SplitArgument{1}{+}}O{N/A+N/A} >{\SplitArgument{1}{+}}O{N/A+N/A} >{\SplitArgument{1}{+}}O{N/A+N/A}}{%
	\resultsCard{#1}{#2}{
		\node[anchor=north east,xshift=3mm] at (4, 1.5)
		{{\FirstArg #3} {\scriptsize\color{gray}{$\pm {\SecondArg #3}$}}};			
		\node[anchor=north east,xshift=2mm,scale=.8,opacity=.7] at (4, .8)
		{#4};			
		\begin{scope}[shift={(0,0)}]
			\node[anchor=north west] at (.35, .5)				
			{\resultsCardImg{experiments/HHL/#2.5qubit-1ancilla-CZ.pdf}[33mm]};
			\node at (2.15,-3.1) {{\FirstArg #5} {\scriptsize\color{gray}{$\pm {\SecondArg #5}$}}};
		\end{scope}	
		\begin{scope}[shift={(0,-3.85)}]
			\node[anchor=north west] at (.35, .5)				
			{\resultsCardImg{experiments/HHL/#2.6qubit-1ancilla-CX-1024.pdf}[33mm]};
			\node at (2.15,-3.1) {{\FirstArg #6} {\scriptsize\color{gray}{$\pm {\SecondArg #6}$}}};
		\end{scope}
		\begin{scope}[shift={(0,-7.7)}]
			\node[anchor=north west] at (.35, .5)				
			{\resultsCardImg{experiments/HHL/#2.7qubit-1ancilla-CX-1024.pdf}[33mm]};
			\node at (2.15,-3.1) {{\FirstArg #7} {\scriptsize\color{gray}{$\pm {\SecondArg #7}$}}};
		\end{scope}
	}{11}
}

\enlargethispage{5cm}

\noindent On some devices with more than $5$ qubits, we also ran some larger instances of the HHL benchmark. Note that, in contrast to the results above, the bottom two runs (matrix size $32$ and $64$) were obtained using only $1024$ shots.
\\[1cm]

\noindent
\HHLSmallResultsCardNext{-}{simulation}{\calcSc{0.015+0.046+0.069}+\calcSc{0.021+0.056+0.080}}{Mar 2021}[0.015+0.021][0.046+0.056][0.069+0.080]
\HHLSmallResultsCardNext{IonQ/Amazon}{IonQ}{\calcSc{0.191+0.362+0.448}+\calcSc{0.022+0.078+0.111}}{Feb \& Apr 2021}[0.191+0.022][0.362+0.078][0.448+0.111]
\HHLSmallResultsCardNext{IBM}{casablanca}{N/A+N/A}{Mar 2021}[0.584+0.023][0.767+0.080][N/A+N/A]
\HHLSmallResultsCardNext{IBM}{melbourne}{\calcSc{0.581+0.590+0.738}+\calcSc{0.024+0.086+0.114}}{Mar 2021}[0.581+0.024][0.590+0.086][0.738+0.114]
\clearpage
\subsubsection{Platonic Fractals}
\DeclareDocumentCommand{\platonicResultsCard}{
    m m
    m m
    o o o
}{%
\resultsCard{#1}{#2}{
    \node[anchor=north west] at (0, .5) {\resultsCardImg{experiments/platonic/#2.1.pdf}};
    \IfValueTF{#5}{\ScoreTwoLinesSmall{0.1, .1}{#5}}{};
    \node[anchor=north west] at (0, -3) {\resultsCardImg{experiments/platonic/#2.2.pdf}};
    \IfValueTF{#6}{\ScoreTwoLinesSmall{0.1, -3.4}{#6}}{};
    \node[anchor=north west] at (0, -6.5) {\resultsCardImg{experiments/platonic/#2.3.pdf}};
    \IfValueTF{#7}{\ScoreTwoLinesSmall{0.1, -6.9}{#7}}{};
    \ScoreTwoLines{4.3, 1.4}{#3}
    \node[anchor=south west,align=left,opacity=.7] at (0, -11) {#4};
}{11}
}

\DeclareDocumentCommand{\platonicSmallResultsCard}{
	m m
	m m
	o o
}{%
	\resultsCard{#1}{#2}{
		\node[anchor=north west] at (0, .5) {\resultsCardImg{experiments/platonic/#2.1.pdf}};
		\IfValueTF{#5}{\ScoreTwoLinesSmall{0.1, .1}{#5}}{};
		\node[anchor=north west] at (0, -3) {\resultsCardImg{experiments/platonic/#2.2.pdf}};
		\IfValueTF{#6}{\ScoreTwoLinesSmall{0.1, -3.4}{#6}}{};
		\ScoreTwoLines{4.3, 1.4}{#3}
		\node[anchor=south west,align=left,opacity=.7] at (0, -7.5) {#4};
	}{7.5}
}

\enlargethispage{5cm}
\noindent
\platonicResultsCard{-}{simulation}{0.01+0.01}{Nov 2020}[0.004+0.003][0.007+0.008][0.014+0.023]
\platonicResultsCard{IBM}{athens}{0.12+0.06}{Oct 2020}[0.071+0.037][0.095+0.065][0.197+0.169]
\platonicResultsCard{IBM}{belem}{0.17+0.08}{Mar 2021}[0.154+0.063][0.174+0.077][0.281+0.220]
\platonicResultsCard{IBM}{bogota}{0.16+0.07}{Mar 2021}[0.162+0.056][0.154+0.084][0.202+0.188]
\\[-5mm]

\noindent
\platonicResultsCard{IBM}{burlington}{0.30+0.10}{Dec 2019}[0.172+0.071][0.350+0.147][0.378+0.242]
\platonicResultsCard{IBM}{casablanca}{0.20+0.08}{Dec 2019}[0.144+0.071][0.149+0.096][0.304+0.216]
\platonicResultsCard{IBM}{essex}{0.24+0.08}{Dec 2019}[0.255+0.114][0.221+0.104][0.249+0.172]
\platonicResultsCard{IBM}{ibmqx2}{0.21+0.07}{Dec 2019}[0.226+0.077][0.159+0.083][0.240+0.191]
\\[-5mm]

\noindent
\platonicResultsCard{IBM}{london}{0.36+0.14}{Dec 2019}[0.143+0.075][0.435+0.301][0.503+0.269]
\platonicResultsCard{IBM}{melbourne}{0.24+0.09}{Dec 2019}[0.124+0.057][0.197+0.086][0.412+0.234]
\platonicResultsCard{IBM}{ourense}{0.18+0.09}{Dec 2019}[0.152+0.142][0.134+0.099][0.239+0.191]
\platonicResultsCard{IBM}{quito}{0.19+0.06}{Mar 2021}[0.225+0.090][0.159+0.088][0.171+0.158]
\\[-5mm]
\enlargethispage{5cm}

\noindent
\platonicResultsCard{IBM}{rome}{0.16+0.05}{Mar 2021}[0.148+0.064][0.177+0.091][0.209+0.200]
\platonicResultsCard{IBM}{santiago}{0.19+0.09}{Sep 2020}[0.169+0.099][0.148+0.112][0.260+0.224]
\platonicResultsCard{IBM}{valencia}{0.17+0.08}{Sep 2020}[0.094+0.039][0.186+0.109][0.232+0.223]
\platonicResultsCard{IBM}{vigo}{0.18+0.08}{Dec 2019}[0.096+0.067][0.221+0.152][0.211+0.167]
\\[-5mm]

\noindent
\platonicSmallResultsCard{Rigetti}{Aspen-8}{?+}{Oct 2020}[0.529+0.370][0.766+0.454]

\clearpage
\restoregeometry
\pagestyle{plain}

\begin{figure}[!t]
    \centering
    \includegraphics[width=13cm]{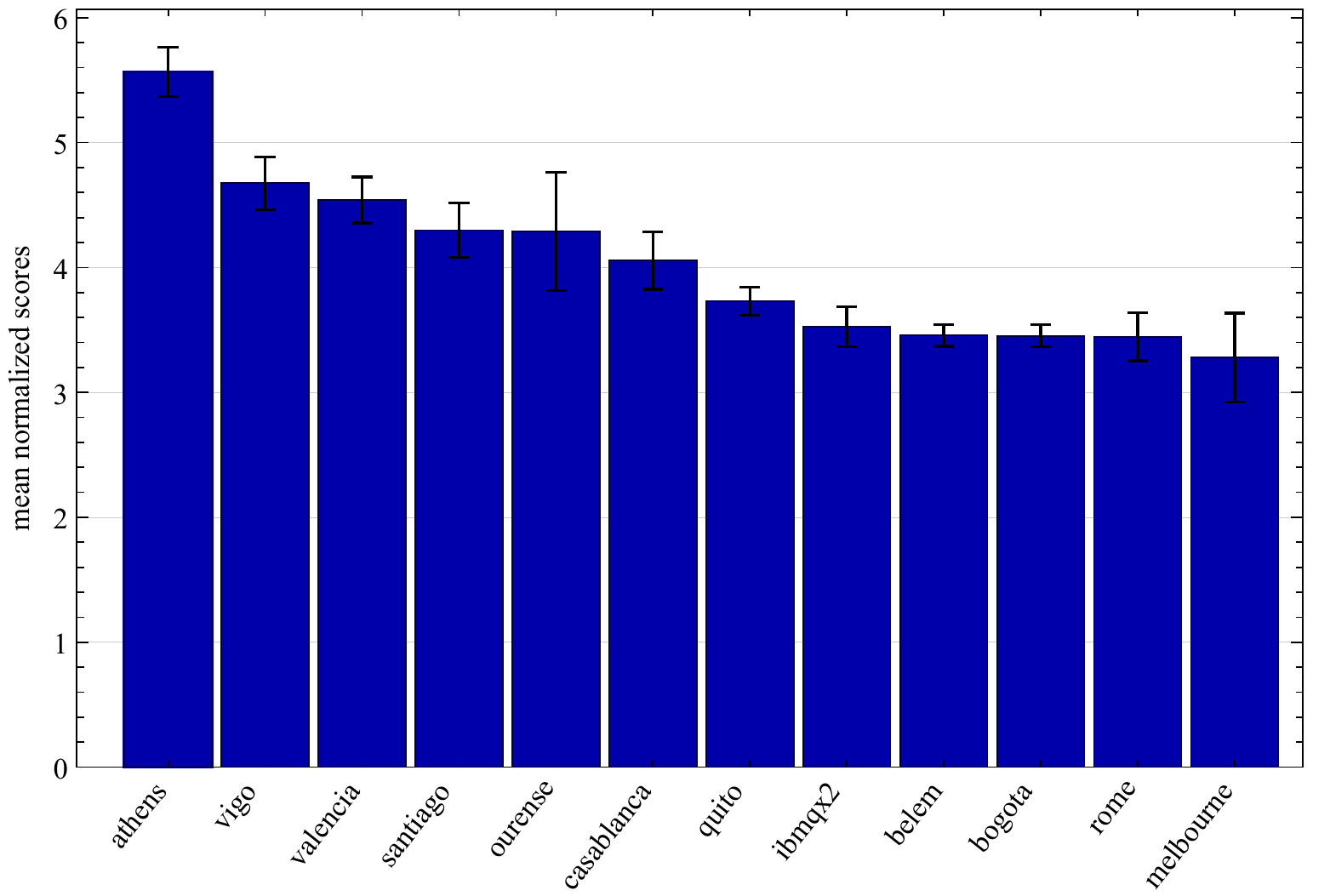}
    \caption{Mean scores for those devices where we ran all six benchmarks, calculated as described in \cref{sec:meanscores}.
    The IonQ and Aspen devices, as well as a few IBM devices, do not have all benchmarks run; as such we do not include it in the comparison.
    }
    \label{fig:meanscores}
\end{figure}

\subsection{Mean Scores}\label{sec:meanscores}
To collate the various test results into a single number, we propose to first normalize the scores (such that they lie within the interval $[0,1]$, corresponding to a factor of $1/2$ for the Platonic Fractals, $1/\sqrt2$ for Line Drawing, and $(1.5-\text{score})/\text{score}$ for the Bell test), take their mean, and then their inverse.
This roughly corresponds to the intuition that if a device is ``half as noisy'', the individual benchmark scores should, roughly, grow ``twice as close'' to their optimum (i.e.\ 0 for all but the Bell test; for the latter 1.5), and hence this inverted mean score yields a number higher by a factor of two.
Taking the inverse of the mean instead of the mean of the inverses also means that the worst benchmarks dominates the final result; indicating that if a device is truly better, then it cannot just simply be optimized for a single type of circuits.

Naturally, in case not all benchmarks have been run (such as in the case of the IonQ device, which we only benchmarked with the HHL test), the mean score has to be taken with caution, so we do not include it in the comparison.
Nonetheless, we can establish an approximate ranking of the tested devices, as shown in \cref{fig:meanscores}.
We emphasize that these overall scores do not necessarily reflect the performance of each device for all tasks.

\begin{figure}[!t]
    \hspace*{-1.5cm}
    \includegraphics[width=18.5cm]{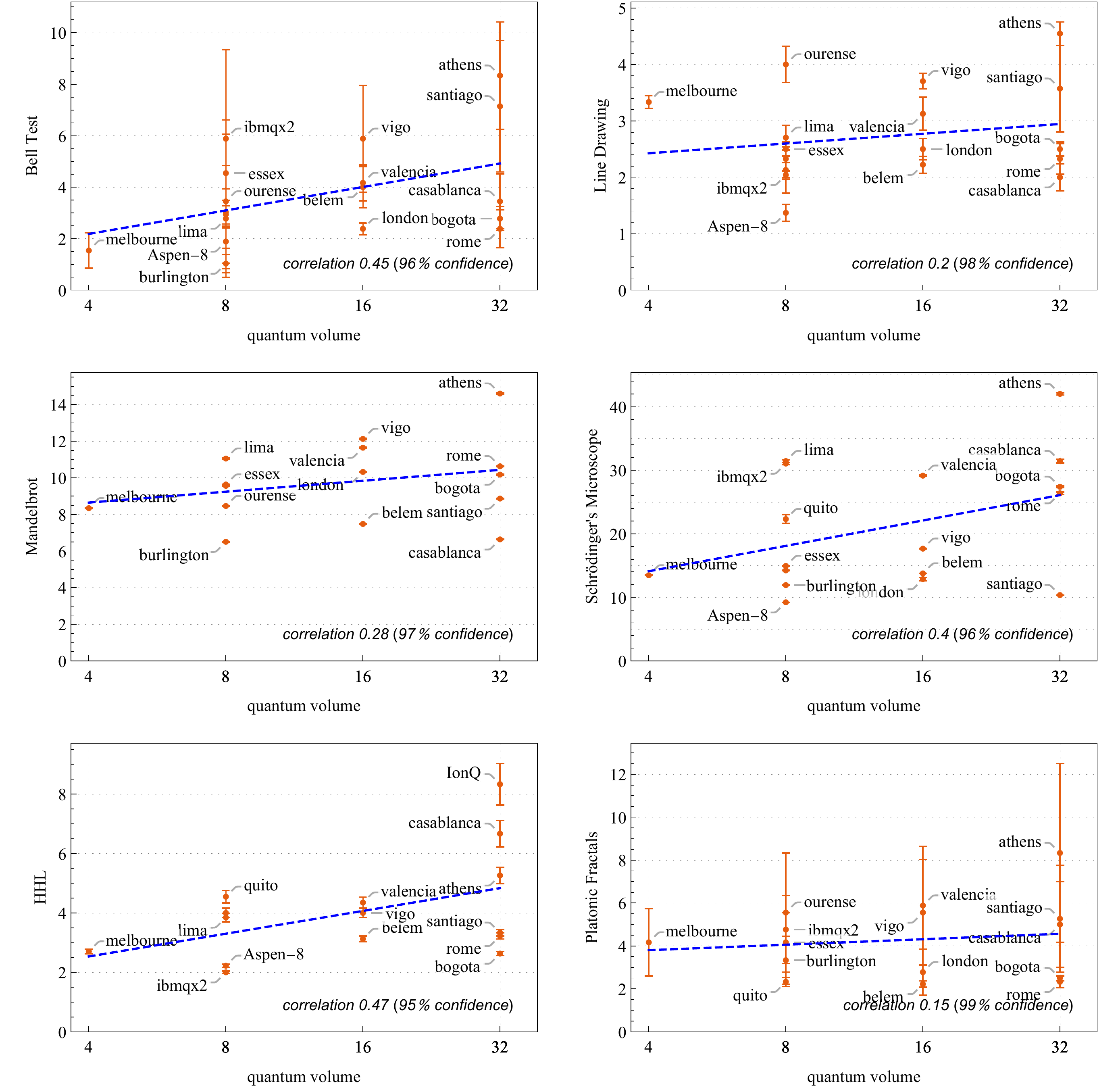}
    \caption{Individual benchmark scores vs.\ quantum volume. The individual correlation coefficients and confidences are given at the bottom of each plot.
    The HHL, Mandelbrot and Schrödinger's Microscope tests and Bell Test correlate the most with quantum volume; Line Drawing and Platonic Fractals the least.}
    \label{fig:qv-all}
\end{figure}

\begin{figure}[!t]
    \centering
    \includegraphics[width=12cm]{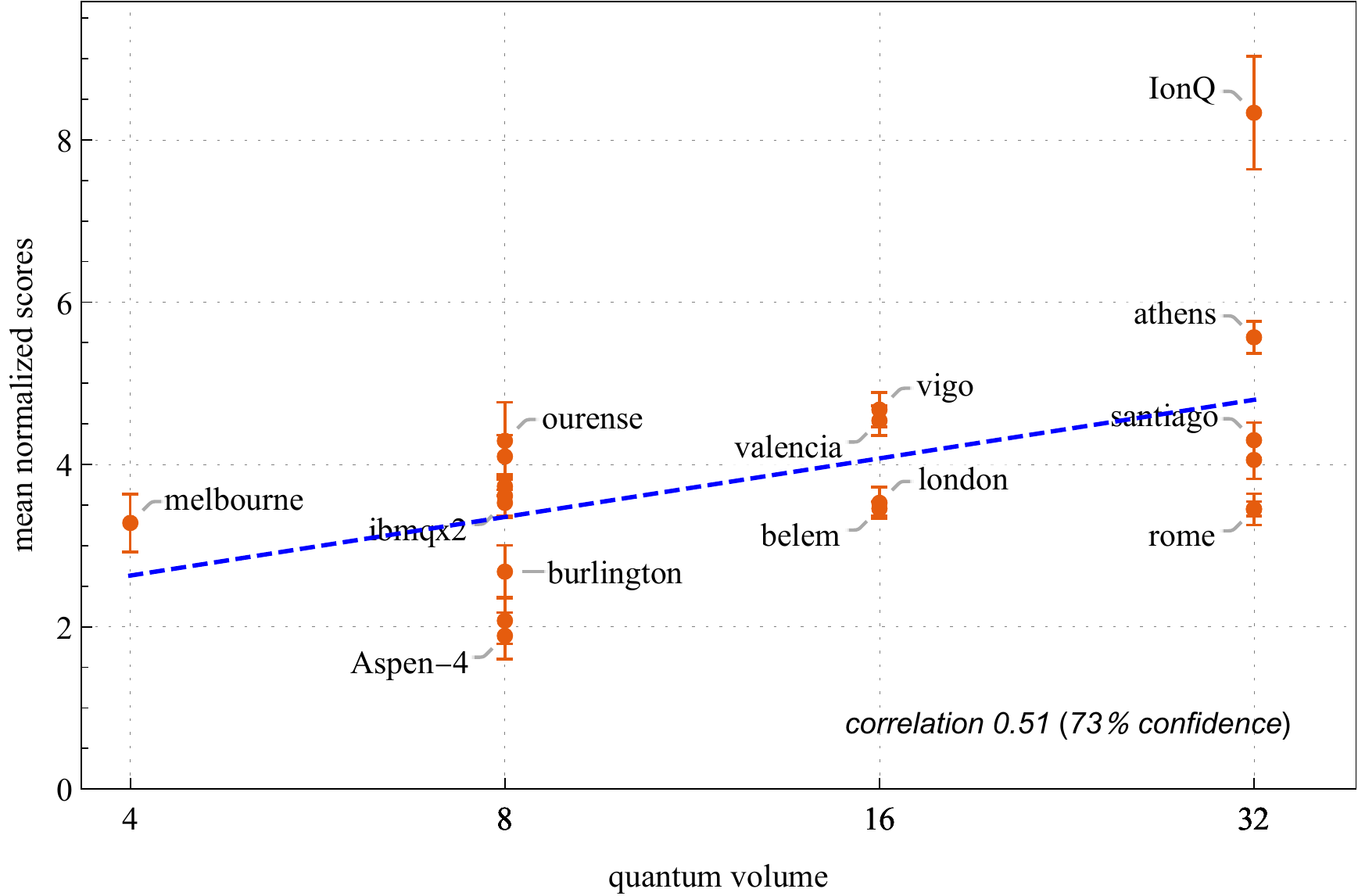}
    \caption{Mean benchmark scores (higher is better) vs.\ quantum volume (higher is better); for the Bell test, a score value of $1.5-$original score was used. With equal weighting, a correlation coefficient of $0.47$ can be extracted with confidence $56\%$.
    As emphasized in \cref{sec:meanscores}, not all devices were tested with all six benchmarks; so the relative scores in this plot should be interpreted with caution.}
    \label{fig:qv-mean}
\end{figure}

\subsection{Comparison with Quantum Volume}\label{sec:qv}
As outlined in the introduction, quantum volume is one of the proposed generic measures for the capabilities of quantum devices.
How much do the scores from our benchmarks correlate with the quantum volume as provided either by the vendors, or measured independently? To this end we relied on data from  \cite{Karalekas2020} (Aspen-4), as well as \cite{Cross2019,GarcaPrez2020} and the IBM Quantum Experience data for the IBM devices (where we relied on measured quantum volume over the one displayed at the IBM Quantum Experience, e.g.\ for melbourne, which at the time of running the tests had a reported QV of 4, and not 8). Aspen-7 and 8 were estimated to have a quantum volume of $\approx 8$ by their average two-qubit gate fidelities of $\approx 95\%$.

\Cref{fig:qv-mean} demonstrates that our benchmarks (where the scores are transformed as in \cref{sec:meanscores}) correlate well with quantum volume; however, the individual tests fare differently, as shown in \cref{fig:qv-all}.

\section{Discussion and Outlook}
Apart from providing an objective and multi-faceted way of assessing the performance of current quantum hardware, there are two core goals we aspire to achieve with this project.
First of all, we wish to provide a guide to the community and the wider public in understanding the strengths and limitations of current hardware.
In particular, we aim to raise awareness about the types of problems that can---or cannot---realistically be implemented with NISQ devices; and a way to track progress over how devices improve over time.
    
Furthermore, since all our benchmarks have an ``exact'' implementation, where the full state vector is simulated, as well as one on an ideal noise-free device, where only measurement errors contribute to the final score, we hope that our tests can provide insights into how gate-level fidelities and $T_1$/$T_2$-times translate into errors within full quantum circuits, potentially extending the ability for device manufacturers to communicate their device capabilities.
As such, the benchmark suite we wrote might prove useful as a first step for assessing how well a device is tuned up, and to provide the user with a more insightful metric than single gate fidelities can.

\paragraph{Acknowledgements.}
The authors acknowledge IBM for providing access to its quantum devices through IBM Quantum, 
Rigetti for providing access to its Aspen-4, Aspen-7 and Aspen-8 quantum devices through Forest,
and Amazon Web Services for providing access to Rigetti's Aspen-9 and IonQ's quantum devices through Amazon Braket.
The views expressed in this paper are those of the authors, and do not reflect the official policy or position of IBM Quantum, Rigetti or Amazon Web Services. 

J.\,B.\ is grateful for support from the Draper's Research Fellowship at Pembroke College.
A.\,G.\ acknowledges funding provided by Samsung Electronics Co., Ltd., for the project ``The Computational Power of Sampling on Quantum Computers''. Additional support was provided by the Institute for Quantum Information and Matter, an NSF Physics Frontiers Center (NSF Grant PHY-1733907), and by ERC Consolidator Grant QPROGRESS as well as QuantERA project QuantAlgo 680-91-034.
A.\,C.\ would like to thank Farrokh Labib, for insightful discussions about the relationship between closed curves in the complex plane, and their expression in terms of Fourier coefficients.

\clearpage
\printbibliography

\end{document}